\documentclass[useAMS,usenatbib]{mn2e}

\usepackage{amssymb, amsmath, epsfig, url}
\newcommand{\bfk}{{\boldsymbol{k}}}

\newcommand{\bfC}{{\mathbf{C}}}
\newcommand{\matF}{{\mathbf{F}}}

\newcommand{\Mpc}{{\rm Mpc}}

\newcommand{\HI}{H{\sc ~i}}

\newcommand{\HeI}{He{\sc ~i}}
\newcommand{\HeII}{He{\sc ~ii}}
\newcommand{\HeIII}{He{\sc ~iii}}

\begin{document}

\title[Temperature Fluctuations in the Ly$\alpha$ Forest]{The Signatures of Large-scale Temperature and Intensity Fluctuations in the Lyman-$\alpha$ Forest}

\author[M. McQuinn et al.]{\parbox[t]{\textwidth}{Matthew McQuinn$^1$\thanks{mmcquinn@berkeley.edu}, Lars Hernquist$^{2}$, Adam Lidz$^{3}$, Matias Zaldarriaga$^{4}$}\\ \\
$^{1}$ Department of Astronomy, University of California, Berkeley, CA 94720, USA\\ 
$^{2}$ Harvard-Smithsonian Center
for Astrophysics, 60 Garden St., Cambridge, MA 02138, USA\\ 
$^3$ Department of Physics and Astronomy, University of Pennsylvania; Philadelphia, PA 19104, USA\\
$^{4}$ Institute for Advanced Study, 1 Einstein Drive, Princeton, NJ 08540, USA\\} 

\pubyear{2009} \volume{000} \pagerange{1}

\maketitle\label{firstpage}

\begin{abstract}

It appears inevitable that reionization processes would have produced
large-scale temperature fluctuations in the intergalactic medium.
Using toy temperature models and detailed heating histories from
cosmological simulations of \HeII\ reionization, we study the
consequences of inhomogeneous heating for the Ly$\alpha$ forest.
The impact of temperature fluctuations in physically
well-motivated models can be surprisingly subtle.  In fact, we show
that temperature fluctuations at the level predicted by our
reionization simulations do not give rise to detectable signatures in
the types of statistics that have been employed previously.  However,
because of the aliasing of small-scale density power to larger scale
modes in the line-of-sight Ly$\alpha$ forest power spectrum, earlier
analyses were not sensitive to $3$D modes with $\gtrsim 30~$comoving Mpc wavelengths -- scales
 where temperature fluctuations are likely to be
relatively largest.  The ongoing Baryon Oscillation Spectroscopic
Survey (BOSS) aims to measure the $3$D power spectrum of the
Ly$\alpha$ forest, $P_F$, from a large sample of quasars in order to
avoid this aliasing.  We find that physically motivated temperature
models can alter $P_F$ at an order unity level at $k \lesssim
0.1~$comoving Mpc$^{-1}$, a magnitude that should be easily detectable
with BOSS.  Fluctuations in the intensity of the
ultraviolet background can also alter $P_F$ significantly.  These
signatures will make it possible for BOSS to study the thermal impact
of \HeII\ reionization at $2 < z < 3$ and to constrain models for the
sources of the ionizing background.  Future spectroscopic surveys
could extend this measurement to even higher redshifts,
potentially detecting the thermal imprint of hydrogen reionization.

%In addition, we show that a particular 3-point statistic that correlates large-scale flux with small scale power could be sensitive to inhomogeneous heating when applied to a sample of $\approx 10$ high-resolution spectra. 

\end{abstract}


\section{Introduction}

The temperature of the intergalactic medium (IGM) is largely
determined by how and when the cosmic hydrogen and helium were
reionized.  Measurements of the mean temperature of the IGM at
redshifts $2 <z< 6$ appear to be consistent with the paradigm that
hydrogen was reionized at $z\sim 10$ and helium was doubly ionized at
$z\sim 3$ \citep{2000ApJ...534...41R, 2000MNRAS.318..817S,
zaldarriaga01c, mcdonald01b, hui03, lidz09, bolton10, becker10}.
However, not all of these studies agree on the trends nor on their
interpretation.
  
In addition to raising the mean temperature of the IGM, reionization
processes would have heated the intergalactic gas inhomogeneously.
Once imprinted, an intergalactic temperature fluctuation would fade
away over roughly a Hubble time \citep{hui03}.  Inhomogeneities in the
temperature of the IGM alter the Ly$\alpha$ forest absorption because
the ionization state of hydrogen depends on the temperature, because the
gas distribution on $\lesssim 100$~kpc scales is smoothed by thermal
pressure, and because of thermal broadening of the absorption features.
However, the vast majority of studies of the Ly$\alpha$ forest have
assumed that there is an approximately power-law relationship between
temperature and density.  Our study investigates the detectability of
realistic models for temperature inhomogeneities.
  
The level of intergalactic temperature fluctuations has been
constrained by previous analyses of the forest.  
\citet{2002MNRAS.332..367T} and \citet{lidz09} 
placed upper limits on the allowed level of these fluctuations in the
context of toy inhomogeneous heating models.  In addition, the standard model
for the Ly$\alpha$ forest has been successful at explaining the
statistical properties of this absorption \citep{miralda96, hernquist96, katz96, dave99, mcdonald05b}.  This picture posits that
the fluctuations in the $2 \lesssim z \lesssim 5$ forest were driven
primarily by density inhomogeneities, that the gas temperature can be
approximated as a power law in density, and that the photoionization
rate was nearly spatially invariant.  The success of this model
implies that deviations from it cannot be large.  %In fact, \citet{mcdonald05} argued that the standard approximations are adequate for the precision of past measurements.  However, these authors focused on a single statistic, the line-of-sight power spectrum.  In what follows, we show that the impact of temperature and intensity fluctuations can be significantly larger in the $3$D Ly$\alpha$ forest power spectrum, a statistic that upcoming surveys aim to measure.

 %(Although, the preferred value of $\sigma_8$ from analyzes of the forest is generally higher than of other cosmological probes \citealt{}.)
  
Several theoretical studies of the Ly$\alpha$ forest have examined the
impact of temperature inhomogeneities on various statistics.  \citet{lai06}
investigated the effect of temperature fluctuations on the Ly$\alpha$
forest line-of-sight power spectrum.   Surprisingly, they found that models with ${\cal O}(1)$ fluctuations in the temperature ($\Delta T/T \sim 1$) that correlated over
tens of comoving Mpc altered this statistic at \emph{only} the few
percent-level for $k < 5~$comoving~Mpc$^{-1}$ compared to models with power-law temperature-density relations.  

Other statistics have been found to be more sensitive to such inhomogeneities.  \citet{meiksin00},
\citet{theuns00}, and \citet{zaldarriaga02} advocated the use of
wavelets filters as a means to spatially identify temperature inhomogeneities,
and they showed that wavelets are sensitive to $10$s of
comoving Mpc fluctuations with $\Delta T/T \approx 1$ when applied to $\approx 10$ high-resolution Ly$\alpha$ forest spectra.   \citet{lee10} found that threshold clustering functions (a popular statistic in the material sciences) could discriminate between different allowed thermal states.  \citet{fang04} argued that a three-point statistic which correlates
the large-scale flux with the small-scale power could place a strong
constraint on deviations from a power-law temperature-density relation.
Lastly, \citet{white10} showed that large temperature fluctuations could have
a significant effect on two- and three-point functions estimated from
correlating multiple Ly$\alpha$ forest sightlines.

Here, we quantify the impact of $10$s of comoving Mpc temperature
fluctuations on a diverse set of Ly$\alpha$ forest statistics.  We use both
toy models as well as the \HeII\ reionization simulations presented in
\citet{mcquinn09} to understand these effects.  Section~\ref{sec:background} discusses the different nonstandard contributions
to fluctuations in the forest, concentrating on inhomogeneities in the
temperature.  Section~\ref{sec:methodology} describes the methods of
analysis used in our study.  Section~\ref{sec:statistics} (and
Appendix~A) quantifies the impact of models for temperature
inhomogeneities on different Ly$\alpha$ forest statistics.
Section~\ref{sec:Gamma} discusses how fluctuations in the ionizing
background could also alter Ly$\alpha$ forest statistics.  

We adopt a flat $\Lambda$CDM cosmological model consistent with the
most recent cosmological constraints \citep{komatsu10}, and we
henceforth will use ``Mpc'' as shorthand for ``comoving Mpc.''  All of
our Ly$\alpha$ forest calculations normalize to the mean flux, $\langle F \rangle$, values of \citet{faucher07}, which measured $\langle F \rangle = 0.69$
at $z=3$ and $\langle F \rangle = 0.39$ at $z=4$.
% Appendix A develops a toy model to understand the evolution of the IGM temperature seen in the simulations of \citet{mcquinn09}. 

%This study uses an eclectic set of simulations to calculate the desired statistics, using hte $35~$Mpc, $2\times 512^3$ cosmological hydrodynamic simulations presented in \citet{lidz09}, several $200-500~$Mpc dark matter simulations that were post processes with a radiative transfer to model \HeII\ reionization in \citet{mcquinn09}, and the suite of $750~$Mpc dark matter simulations of the forest presented in \citet{white09}.  We use the simulation that is best suited for quantifying the chosen statistic.  While the three sets of simulations use slightly different parameter values for a $\Lambda$CDM cosmology, all three are consistent with the most recent cosmological constraints on this model \citep{komatsu10}.

\section{Background}
\label{sec:background}

In the Sobolev approximation, the optical depth for a photon to be absorbed as it redshifts across the Ly$\alpha$ resonance of hydrogen is
\begin{equation}
\tau_{\rm Ly\alpha} \approx 1.1 \, \Delta_b^2 \; T_4^{-0.7}\;  \Gamma_{-12}^{-1} \; \left(\frac{1 +z}{4}\right)^{{9}/{2}} \; \frac{H(z) /(1+z)}{dv/dx},
\label{eqn:tauHI_GP}
\end{equation}  
where $\Delta_b$ is the gas density in units of the cosmic mean, $T_4$ is the temperature in units of $10^4$~K, $H(z)$ is the Hubble parameter, and $dv/dx$ is the line-of-sight velocity gradient (which is equal to $H(z)/(1+z)$ in the absence of peculiar velocities).  Equation (\ref{eqn:tauHI_GP}) assumes that the hydrogen is in photoionization equilibrium with a photoionization 
rate of $\Gamma_{-12}$, expressed in units of $10^{-12}\,$s$^{-1}$.  Measurements find $\Gamma_{-12} \approx 1$ (e.g., \citealt{faucher08}).  %Photoionization equilibrium is an excellent assumption because the equilibration time is $\Gamma_{-12}^{-1}$, which is much smaller than the Hubble,  dynamical, and cooling times for gas at relevant temperatures and densities.  
The $T_4^{-0.7}$ factor arises because of the temperature 
dependence of the \HI\ fraction in photoionization equilibrium, and many of the effects studied here derive from this dependence.  

The amount of Ly$\alpha$ absorption is determined by inhomogeneities in the gas density, the peculiar velocity field, the photoionization rate, and the gas temperature.  The first two sources are the standard contributions that were included in most previous studies and that are primarily responsible for the statistical properties of the Ly$\alpha$ forest.  Almost all Ly$\alpha$ forest studies also have assumed that the temperature follows a power-law relation in density parametrized as $T(\Delta_b) = T_0 \, \Delta^{\gamma - 1}$ (or they have effectively assumed a power-law relation by using simulations with a spatially uniform ionizing background).  This study focusses on the impact of temperature fluctuations around $T(\Delta_b)$, and it also considers fluctuations in the photoionization rate.  In addition to modulating the amplitude of $\tau_{\rm Ly\alpha}$, inhomogeneous heating alters $10$s of km~s$^{-1}$ absorption features via its impact on the gas pressure and on the thermal widths of absorption lines.
%\footnote{Temperature inhomogeneities also affect the absorption by other Lyman-series lines in the same way as Ly$\alpha$, but we concentrate on the Ly$\alpha$ here because it is sensitive to the lowest densities where it is easiest to disentangle the effect of \HeII\ reionization and because the higher Lyman series transitions are contaminated by foreground Ly$\alpha$ absorption.}

Large-scale temperature fluctuations should be present in the IGM as a relic of reionization processes.  The reionization of hydrogen and helium were spatially inhomogeneous, with some regions ionized earlier and some later depending on their proximity to the ionizing sources.   Models for these processes predict order unity fluctuations in the ionization fraction of the species being reionized on $\sim 10$~Mpc scales and accompanying order unity fluctuations in the temperature (e.g., \citealt{trac08, mcquinn09}).  Temperature fluctuations should have been imprinted 
on the IGM during reionization because: (1) different regions in the IGM would have been ionized by different energy photons, and (2) a gas element's instantaneous temperature depends on its ionization history.  Once a typical element was reionized and heated, it would have subsequently cooled adiabatically owing to the expansion of the Universe (with Compton cooling off of the CMB and line cooling playing a minor role).  This cooling proceeded until the gas element reached the temperature at which heating from ionizations of the residual bound electrons balanced the cooling from expansion \citep{miralda94, hui97, theunsschaye02}.  At this point, the temperature difference with initially cooler elements would have been erased.  At $z <6$, it takes on the order of a Hubble time to erase temperature fluctuations.\footnote{To quantify this, we have solved for the evolution of $T_0$ as in \citet{hui97} after hydrogen and helium reionization, assuming a flat specific intensity at the \HI, \HeI, and \HeII\ ionization edges (but the results depend weakly on this assumption).  If half of the gas were at $T_0= 15,000~$K and half at $T_0= 25,000~$K at $z=3$, by $z=2$ the amplitude of fluctuations in $T_0$ would have been $70\%$ of what they had been, half at $17,000~$K and half at $12,000~$K.  If instead half the IGM were at $T_0= 15,000~$K and half at $T_0= 25,000~$K at $z=4$, the amplitude of temperature fluctuations at $z=2$ would have been reduced to $40\%$ of its initial value.}

The Ly$\alpha$ forest indicates that hydrogen was likely reionized at $z >6$.  For any physical radiation spectrum 
that reionized the hydrogen, the intergalactic helium would have been 
at least singly ionized simultaneously with the hydrogen.  At $z < 4$, the relic temperature fluctuations from $z > 6$ reionization processes would have largely faded away \citep{hui97, trac08}.  However, a hard source of ionizing photons 
is required to doubly ionize helium (normal stars cannot do it).  The current paradigm is that the intergalactic helium was doubly ionized by the radiation from quasars at $z \sim 3$ \citep{davidsen96, 2000MNRAS.318..817S, agafonova07, mcquinnGP, becker10, shull10, furlanettodixon, dixonfurlanetto}.   Models predict that at $z\approx 3$ \HeII\ reionization would have imprinted large inhomogeneities in the temperature, much larger than those remaining from hydrogen reionization \citep{furlanetto07a, mcquinn09}.  

%Observations of the quasar luminosity function suggest that quasars near $L_*$ in the Schechter function would have dominated the \HeII-ionizing emissivity at these times \citep{hopkins07a}.  Such quasars should have generated large-scale temperature inhomogeneities because of their $10-100~\Mpc$ \HeIII\ regions \citep{furlanetto07a, mcquinn09}.
\subsection{Simulations of the IGM Temperature}

Using radiative transfer simulations, \citet{mcquinn09} made
predictions for the level of intergalactic temperature fluctuations
expected during \HeII\ reionization.  These simulations are employed
extensively in our new study.  In these simulations, the first regions
in which the \HeII\ was reionized were closest to quasars and,
therefore, were ionized by softer photons and heated by less than
$10^4~$K.  These regions subsequently cooled with the expansion of the
Universe.  Whereas, the last regions to be ionized were more
significantly heated by ionizations from a hardened radiation
background.  Thus, the temporal extent of this process resulted in large fluctuations in the intergalactic temperature.  See \citet{abel99} for discussion of relevant radiative transfer effects.

\begin{figure}
\begin{center}
{\epsfig{file=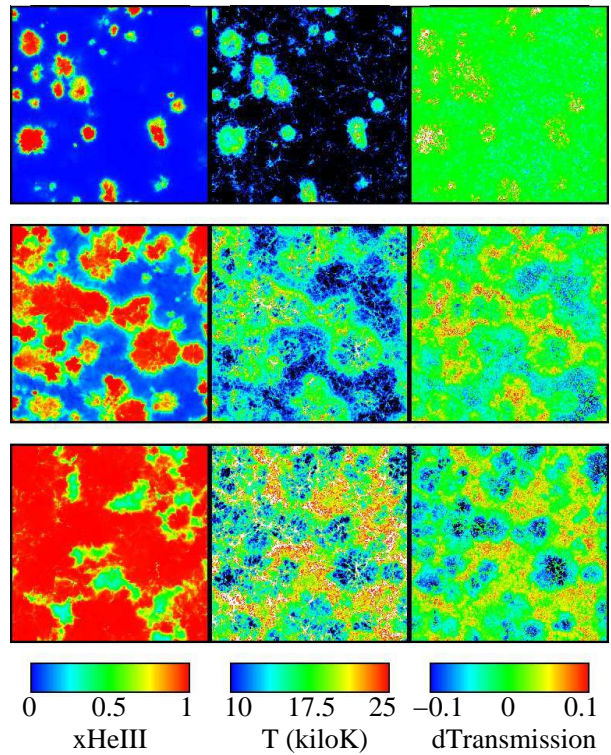, width=8cm}}
\end{center}
\caption{Slices through simulation B1 from \citet{mcquinn09}.  The left panels show the \HeIII\ fraction, the middle panels the temperature, and the right panels deviations in the transmission relative to the case without temperature fluctuations but with the same mean $T(\Delta_b)$.   Each panel is $430 \, \Mpc \times 430 \, \Mpc \times 8 \, \Mpc$ in size, with (from top to bottom) volume-averaged \HeII\ fractions of $\bar{x}_{\rm HeII} = 0.1, 0.5,$ and $0.9$ at $z = 5, ~4$, and $3.5$, respectively. 
\label{fig:big_box}}
\end{figure}

\begin{figure}
\begin{center}
{\epsfig{file=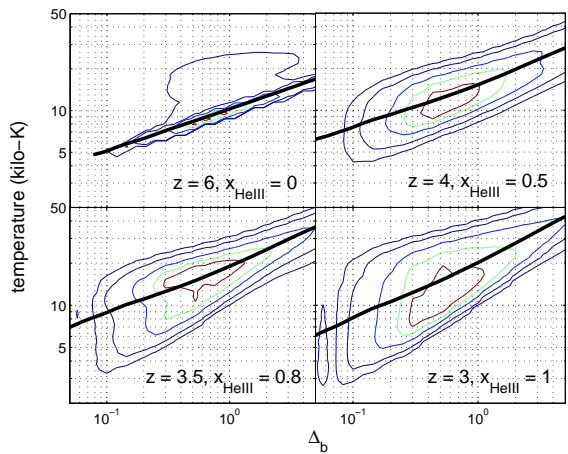, width=8.5cm}}
\end{center}
\caption{Distribution of gas elements in the $T$-$\Delta_b$ plane from the D1 simulation in \citet{mcquinn09}, a smaller box version of the simulation featured in Figure \ref{fig:big_box}.  The thick solid
curves are the mean $T(\Delta_b)$, and the contours enclose $33, 67,
90, 99$, and $99.9\%$ of the grid cells.  These contours enclose the regions that have the highest density of gas elements in $\log T$ - 
$\log \Delta_b$.  \label{fig:Tdeltadist}}
\end{figure}

Figure \ref{fig:big_box} shows slices through three snapshots of a
$430~$Mpc \HeII\ reionization simulation from \citet{mcquinn09}.  The
left panels are the \HeIII\ fraction, the middle panels are the
temperature, and the right panels are the change in the Ly$\alpha$
forest transmission relative to the case with the same mean relation
between $T$ and $\Delta_b$, but without temperature fluctuations.  By
the end of \HeII\ reionization in this simulation, the temperature at
the cosmic mean density fluctuated between $10$ and $30~$kiloK,
resulting in $\approx 10\%$ fluctuations in the Ly$\alpha$ forest
transmission.%\footnote{\citet{mcquinn09} found that the character of temperature fluctuations in these simulations was not affected significantly by how they parameterized their subgrid model for the effect of unresolved \HeII\ Lyman-limit systems and how the thermal state of the IGM was initialized in the simulation (at least at $z<4$).  They found that the amount of heating was most sensitive to the spectral index adopted for quasars.  However, the level of temperature fluctuations increased only marginally in the runs with a harder spectrum.  We do not use the hotter simulations because significantly higher temperatures than the fiducial simulation are inconsistent with the data (Section \ref{sec:statistics}).  
%The correlation length of the simulated temperature fluctuations is sensitive to the lifetime of quasars.  The quasars in the simulations employed here have a lifetime of $\approx 10^8~$yr.  See \citet{mcquinn09} for the comparison of simulations with different quasar lifetimes and light curves.}

We briefly summarize the characteristics of the temperature fluctuations in these simulations, but see \citet{mcquinn09} for additional details.
The solid curves in Figure \ref{fig:Tdeltadist} show the mean temperature as a function of $\Delta_b$ at different times in a higher resolution version of the simulation in Figure \ref{fig:big_box}.    All of these curves can be approximated by a power-law temperature density-relation with $\langle \gamma - 1 \rangle \approx 0.35$.  However, there is significant scatter around this relation, with the scatter increasing as \HeII\ reionization proceeds.  
Moreover, the scatter is larger at lower $\Delta_b$, a property that affects the observability of these fluctuations (Section \ref{sec:statistics}).   The contours in Figure \ref{fig:Tdeltadist} enclose $33, ~67, ~90,~ 99$, and $99.9\%$ of the grid cells.

\begin{figure}
\begin{center}
\rotatebox{-90}{\epsfig{file=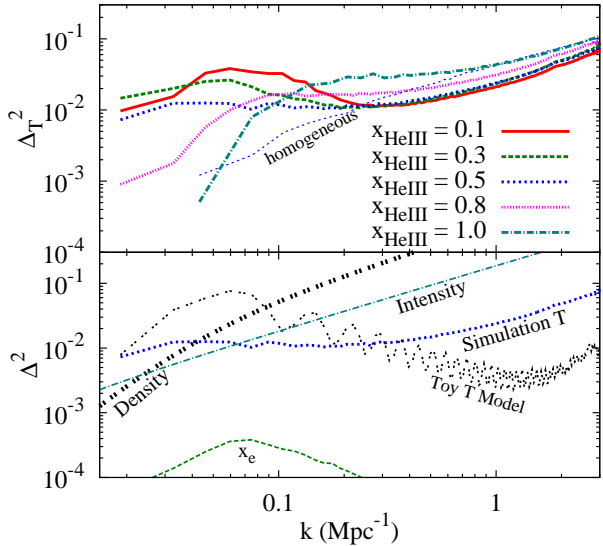, height=11cm}}
\end{center}
\caption{Top Panel:  Magnitude of temperature fluctuations, quantified in terms of the dimensionless power spectrum, $\Delta_T^2 \equiv k^3 P_{T}/ 2\pi^2$, where $P_{T}$ is the power spectrum of $\delta_T \equiv T/\bar{T} - 1$.  The thick curves show $\Delta_T^2$ for different \HeIII\ fractions in the fiducial simulation of \HeII\ reionization.  The thin blue dotted curve labeled ``homogeneous'' assumes $\gamma \approx 1.6$ and $T_0 \approx 10^4~$K. 
Bottom Panel:  $\Delta^2 \equiv k^3 P_{X}/ 2\pi^2$, where $P_{X}$ is the power spectrum of the overdensity in $X$.  The blue dotted curve is for the same temperature model as in the top panel, and the curve labeled ``Toy T Model" is $\Delta_T^2$ for $\gamma - 1 = 0.3$, where half the IGM is filled with uncorrelated $30$~Mpc spheres with $T_0 = 20,000$~K and the other regions have $T_0=10,000$~K.  The dashed green curve labeled $x_e$ is $\Delta^2$ for electron fraction fluctuations, and the curve labeled ``Density'' is this for density fluctuations. % plot located in /o/mmcquinn/programs/fft_routines/fft_float/data_files/HeII_reion_301/fluc_fluc.gnu
\label{fig:fluctuations}}
\end{figure}

The top panel in Figure \ref{fig:fluctuations} plots the dimensionless $3$D power spectrum of temperature fluctuations, calculated from different times during the \HeII\ reionization simulation.  We use the notation $\Delta_X^2 \equiv k^3P_X(k)/2\pi^2$, where $P_X$ is the power spectrum of the overdensity in $X$. The power law-like scaling of $\Delta_T^2$ at $k \gtrsim 1~$Mpc$^{-1}$ owes to the temperature fluctuations being highly correlated with the small-scale density fluctuations because of adiabatic cooling and heating.  The shape of $\Delta_T^2$ at $k \lesssim 1~$Mpc$^{-1}$ owes to the large-scale inhomogeneous heating during \HeII\ reionization (top panel, Figure \ref{fig:fluctuations}).  Fluctuations at the $\sim 10\%$ level are present even for the largest modes in our $430~$Mpc box during the bulk of \HeII\ reionization.  

The bottom panel in Figure \ref{fig:fluctuations} compares $\Delta_T^2$ from the snapshot with $x_{\rm HeII} = 0.5$ (dotted curve) with other relevant sources of fluctuations.   A common toy model for IGM temperature fluctuations takes half of the IGM to be at $10~$kiloK and the other half to be at $20~$kiloK.  The curve labeled ``Toy T Model'' is such a model, in which $30~$Mpc, $20~$kiloK bubbles are placed randomly until they fill half of the volume.  The ringing behavior owes to the artificial top-hat bubble morphology.  The fluctuations in this model are a few times larger than in the simulation.  Thus, a similar level of temperature fluctuations as in the simulations should be more difficult to detect than in this model.  This is an important point because \citet{2002MNRAS.332..367T} and \citet{lidz09} found they could rule out this toy model using a wavelet analysis (Section \ref{ss:wavelet}).  The curve labeled ``Density'' is $k^3 \,P_{\Delta}(k)/2\pi^2$ at $z=3$, where the non-linear density power-spectrum $P_{\Delta}(k)$ is calculated using the \citet{peacock96} fitting function.   The temperature fluctuations in the simulation can be as large as the density fluctuations at the smallest $k$ captured, which suggests that such temperature fluctuations would have a significant effect on the large-scale transmission fluctuations.
%\footnote{Note however that the temperature fluctuations appear to the $0.7$ power in $\tau$ (eqn. \ref{eqn:tauHI_GP}).  If both $\tau \ll 1$ and $\delta T \ll 1$, this leads to a factor of $\sim 2$ smaller fluctuation level.}

\HeII\ reionization also produces additional free electrons, increasing the electron abundance (and $\tau_{\rm Ly\alpha}$) by $8\%$ in \HeIII\ regions.  The power spectrum of the electron fraction from the $\bar{x}_{\rm HeII} = 0.5$ snapshot is represented by the  dashed green curve in the bottom panel in Figure \ref{fig:fluctuations}.  The temperature fluctuations in the simulation are almost an order of magnitude larger than the fluctuations in $x_e$.

An inhomogeneous ultraviolet background (which modulates $\Gamma_{-12}$) is the final potential nonstandard source of transmission fluctuations in the forest and is also the most studied \citep{zuo92, zuo92b, meiksin04, croft04, mcdonald05, furlanettoJfluc, 2009MNRAS.400.1461M, white10}.  The curve labeled ``Intensity" represents an empirically motivated model for intensity fluctuations at $z=3$ (see Section \ref{sec:Gamma}).  Thus, the intensity power can also be comparable to that in the density at the smallest $k$.    %Uncertainties in the quasar luminosity function translate to a factor of several uncertainty in the normalization of this curve.

\section{Methodology}
\label{sec:methodology}
We use two of the \HeII\ reionization simulations presented in \citet{mcquinn09}.  The radiative transfer calculation in these simulations was run on a $256^3$ grid in post-processing using the density field from either a $190$ or $430~$Mpc cosmological $N$-body simulation.  Both simulations employ the same model for the sources.  % and \HeII\ reionization ends $\Delta z \approx 0.3$ later in the $190~$Mpc simulation because it captures better the number of recombinations.  
In addition to the \HeII\ reionization simulations, we use three supplementary simulations, all initialized with different random seeds.  We use a $4000^3$ $N$-body 
simulation described in \citet{white10}.  This simulation has dimensions of $750~$Mpc/h and the dynamics were softened at $100~$kpc/h to approximate pressure smoothing for $\sim 10^4~$K gas.  The other two supplementary simulations are $25~$Mpc/h Gadget-2 runs \citep{springel05} described in 
\citet{lidz09} with either $512^3$ or $1024^3$ dark matter and SPH particles.
The supplementary simulations are used to overcome the resolution and sample variance limitations of the \HeII\ reionization simulations, as well as to include hydrodynamical effects absent in the reionization simulations.   All of these simulations were initialized with cosmological parameters that are consistent with the measurement of \citet{komatsu10}. 

We add fluctuations in the temperature to the supplementary simulations
using both toy models and a method that employs the temperature
information from the \HeII\ reionization simulations.  The latter
method relies on the property that locally the $T$--$\Delta_b$ relation
is well-approximated by a power law even though globally a power law
is a poor approximation.  A local power law holds because neighboring
cells have roughly the same thermal history.  Specifically, this
method obtains $T_0$ and $\gamma$ for a cell in the \HeII\
reionization simulation by fitting this power-law model to $T$ and
$\Delta_b$ in the surrounding $5^3$ cells, and it does this along one
long diagonal skewer that cycles through the \HeII\ reionization
simulation box.  It then uses the resulting $T_0$ and $\gamma$ to
transfer the temperature field to density and velocity skewers
extracted from the supplementary simulations (which we make periodic
with length equal to the box size).

\begin{figure}
\rotatebox{-90}{\epsfig{file=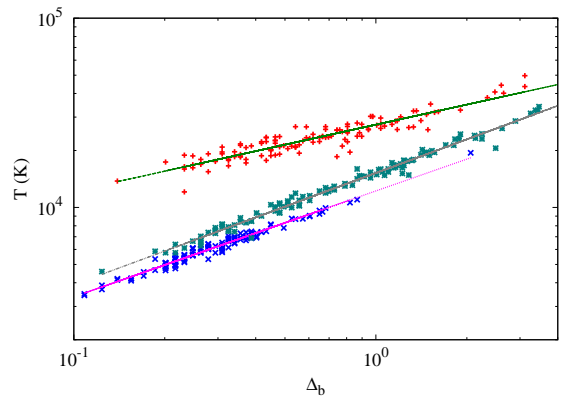, height=8cm}}
\caption{Fits to the local $T$--$\Delta_b$ relation using the $z = 3.0$ snapshot from the $190~$Mpc \HeII\ reionization simulation.  The lines represent the best-fit $T$--$\Delta_b$ relation in $3$ separate, randomly selected regions of $5^3$ cells (corresponding to $3.6$~Mpc on a side).  The markers represent the value of $T$ and $\Delta_b$ in each set of $5^3$ cells.
\label{fig:Tdelta_local}}
\end{figure}

Figure \ref{fig:Tdelta_local} examines the validity of this method.  The three sets of $5^3$ points with different markers represent the values of $T$ and $\Delta_b$ for the cells in three randomly selected regions of size $3.6$~Mpc in the fiducial \HeII\ reionization simulation.  The lines are the best-fit power law to each set of points.  We find that $\Delta T/T \approx 0.1-0.2$ is typical of the error around the best-fit $T-\Delta_b$ relation, which is much smaller than the global dispersion in the temperature.  Thus, the local power-law approximation appears to be justified.  However, this method assumes that the correlations between the large-scale temperature and density modes are unimportant.  This is motivated by the fact that the correlation between temperature and large-scale density is weak in the reionization simulations owing to the rare nature of the sources.  We will elaborate on the applicability of this approximation later.

We will often use our method to add the \HeII\ reionization simulation temperature field at one redshift to a snapshot in the supplementary simulations from another redshift.  This makes possible 
visual comparison of how the temperature fluctuations affect the statistic in question throughout the reionization process without the confusion of cosmological evolution in the density field.  In addition, there is uncertainty in exactly when \HeII\ reionization by quasars would have occurred:  \citet{lidz09} argued from the inferred temperature history that the end was at $z\approx 3.5$.  Whereas, \citet{mcquinnGP} and \citet{shull10} argued that it was not complete until at least $z=2.7$, citing the presence of \HeII\ Ly$\alpha$ Gunn-Peterson troughs down to this redshift.  %The most important determinant of the mean temperature and magnitude of temperature fluctuations in our reionization simulations is the global \HeIII\ fraction.

\section{Statistics}
\label{sec:statistics}

\subsection{Small-scale Line-of-Sight Power Spectrum}
\label{ss:sslos}

The line-of-sight power spectrum of transmission fluctuations is the principle statistic that has been used to derive cosmological constraints from the Ly$\alpha$ forest.  Here we investigate how temperature fluctuations affect this statistic at $k \sim 0.1$~s~km$^{-1}$, and Section \ref{ss:LSC} 
examines larger scales.  Studies of the forest marginalize over $T_0$, $\gamma$, and (in some cases) the reionization redshift to derive cosmological constraints, but the impact of the thermal history has the potential to be more complicated.

Thermal broadening is the most important temperature-dependent effect on the small-scale power spectrum.  Pressure smoothing of the gas can also smooth out the small-scale transmission fluctuations.  However, pressure effects are not included self-consistently in any of the simulations analyzed here.  
%The simulations with gas assume reionization occurs homogeneously at $z\approx10$, and the simulations without it do not account for the thermal history on the distribution of matter.  
There are physical arguments for why pressure smoothing is less important to include than thermal broadening:
\begin{enumerate}
\item \citet{gnedin98} showed analytically that in an expanding universe the scales at which pressure damps the growth of linear modes are a couple times smaller than the Jeans wavelength and smaller than the scale at which thermal broadening erases fluctuations.  Both \citet{gnedin98} and \citet{peeples09a} confirmed with simulations that thermal broadening dominates the exponential damping of the small-scale power in the forest. 

\item The timescale for a typical Ly$\alpha$ forest absorber to relax to equilibrium after a heating event is comparable to the Hubble time \citep{gnedin98}.  During \HeII\ reionization, the additional pressure smoothing from the associated heating would not have had a significant effect on the statistics of the forest if this heating occurred within a redshift interval of $\Delta z \approx 2$ (see Fig. 29 in \citealt{lidz09}). %Furthermore, once a Jeans scale gas element is heated to a temperature $T_1$ from $T_0$, if the subsequent expansion is adiabatic it will cool to a temperature $\approx (T_0 \, T_1)^{1/2}$ once relaxed.

\end{enumerate}
Heating will also induce velocity gradients that broaden the majority of absorption systems \citep{2000MNRAS.315..600T}.  The effect of this broadening mechanism on the Ly$\alpha$ forest power-spectrum has not been quantified in as much detail.  This process is unlikely to lead to an exponential cutoff in the small-scale power like thermal broaden because systems with zero width in redshift-space after including this process can still exist.

Thus, this study ignores the dynamical response of the gas, a common approximation in such analyzes.  However, this may bias the interpretation in this section and that in Section \ref{ss:wavelet} towards favoring higher temperatures.  A treatment that includes this effect would be challenging.  

The shape of small-scale Ly$\alpha$ forest power spectrum is sensitive to the average temperature as well as the temperature distribution. If $f_1$ of the volume is at temperature $T_1$ and the rest is at temperature $T_2$, thermal broadening results in the small-scale power spectrum achieving the limiting form at high $k$ of
\begin{equation}
 P_F^{\rm los}(k) \rightarrow f_1\exp \left(-a \, T_1 \, k^2 \right) + \left(1-f_1 \right) \exp \left(-a \, T_2 \, k^2 \right),
\end{equation}
where $a = 2 k_b/m_p$.
In the limit that $T_2 > T_1$ and $a \, T_2 \, k^2 \gg 1$, the power spectrum is solely determined by the regions with $T_1$.  

An additional consideration is that the small-scale power at $z\sim3$ is dominated by slightly overdense regions because these are the regions that contribute the narrowest absorption lines.  At $z=3$, simple tests show that power on the exponential tail of $P_F^{\rm los}$ is primarily from regions with $\Delta_b \approx 2-3$, and this characteristic density decreases with increasing redshift.  The dispersion in the temperature resulting from an inhomogeneous reionization process should be smaller in overdense regions than in underdense ones.  This trend is evident in the simulations (Fig. \ref{fig:Tdeltadist}), and it acts to reduce the impact of temperature fluctuations on the small-scale shape of $P_F^{\rm los}$.

\begin{figure}
\rotatebox{-90}{\epsfig{file=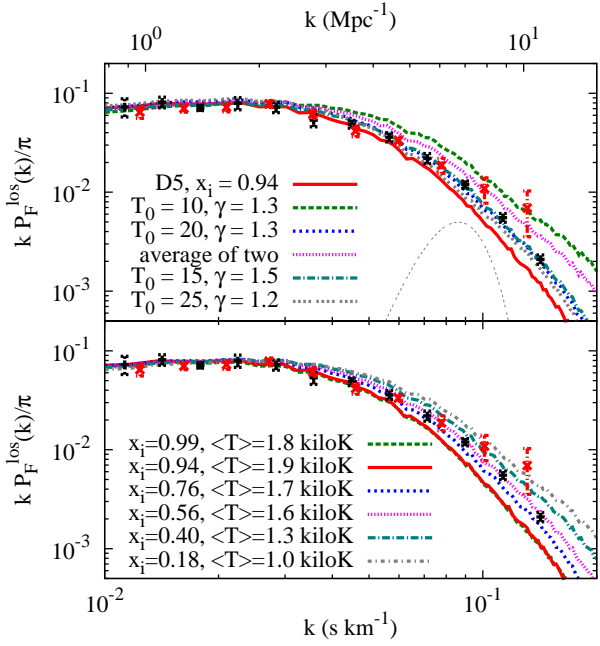, height=9cm}}
\caption{Top panel:  The line-of-sight power spectrum of $\delta_F$ at $z=3$ calculated from the $2\times512^3$ particle $25~$Mpc$/h$ simulation and focussing on the small scale affected by thermal broadening.  The D5 curve is from the fiducial \HeII\ reionization simulation, the other curves are power-law $T-\Delta_b$ models, and the ``average of two'' curve is the average of the $10~$kiloK and $20~$kiloK models with $\gamma = 1.3$.  Bottom panel:  $P_{F}^{\rm los}$ at $z=3$ calculated using the temperature field from different simulation snapshots, labeled by their \HeIII\ fraction ($x_i$ in the plot).  In both panels, the black points with error bars are the measurement of \citet{mcdonald00} and the orange are the measurement of \citet{croft02}.  The temperature field is taken from the snapshots at $z=2.7, 3, 3.3, 3.6, 4$, and $4.5$ in the simulation.  The thin Gaussian curve in the top panel is the wavelet filter used in Appendix A, arbitrarily normalized. 
\label{fig:pk}}
\end{figure}

\begin{figure}
\rotatebox{-90}{\epsfig{file=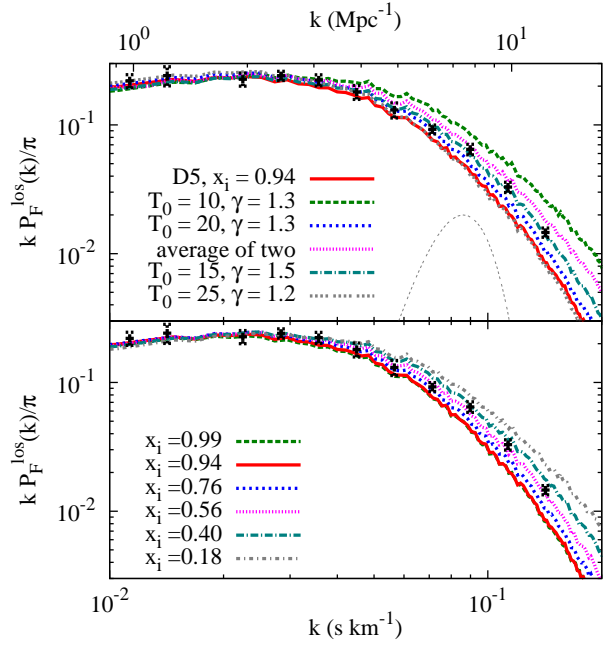, height=9cm}}
\caption{Same as Figure \ref{fig:pk} but for $z=4$ and calculated from the $2\times1024^3$ particle simulation.  See the key in Figure \ref{fig:pk} for the mean temperature of the reionization simulation model curves. 
\label{fig:pk2}}
\end{figure}

Figures \ref{fig:pk} and \ref{fig:pk2} plot the small-scale line-of-sight power spectrum of $\delta_F$ for several temperature models at $z=3$ and $z=4$, respectively.  All curves were calculated from the $25/h$~Mpc Gadget-2 simulations (Section \ref{sec:methodology}), and the points with error bars are the measurements of \citet{mcdonald00} and \citet{croft02}.  In the top panels, the green dashed (blue dotted) curves are for an IGM with $\gamma = 1.3$ and $T_0 = 10~$kiloK ($20~$kiloK).  If $\gamma = 1.3$ and half of the IGM had $T_0 = 10~$kiloK and the other half $T_0 = 20~$kiloK, then the small-scale power spectrum would be given by the magenta dotted curve, which has a slightly flatter shape than the single $T_0$--$\gamma$ models.  The the shape of this curve appears to be inconsistent with the \citet{mcdonald00} data.   However, a toy model for the temperature field that is better motivated by the \HeII\ reionization simulations is half of the IGM at $T_0 = 15~$kiloK with $\gamma = 1.5$ and half at $T_0 = 25~$kiloK with $\gamma =1.2$.  The difference in the small-scale power spectrum between these two states is negligible. (Compare the teal dot-dashed and grey quadruple-dotted curves in the top panels.)  Therefore, the fluctuations in this toy model would not be detectable in $P_F^{\rm los}$.
% because the relative temperature difference is smaller between these models than in the $T_0= 10~$kiloK and $T_0=20~$kiloK case, especially the relative temperature in overdense regions (Figures \ref{fig:pk} and \ref{fig:pk2}).  

The bottom panels in Figures \ref{fig:pk} and \ref{fig:pk2} are the power spectrum of models that use the temperature field of different snapshots from the $190~$Mpc \HeII\ reionization simulation.  The IGM is gradually heated throughout \HeII\ reionization, which results in the small-scale power decreasing with time, or equivalently, with \HeIII\ fraction.  The amount of evolution is sensitive to the simulation's initial conditions, especially with decreasing $\bar{x}_{\rm HeIII}$.  This simulation was initialized with $T_0 = 10~$kiloK and $\gamma = 1.3$ at the starting redshift of $z=6$.  If the simulation were instead initialized with $T_0 = 15~$kiloK, there would have been less evolution.  

At the end of \HeII\ reionization, the simulation temperature reaches values that are slightly hotter than suggested by the $z=3$ measurements of \citet{mcdonald00} and \citet{croft02}.  The $x_{\rm HeIII}=0.94$ curve in Figure \ref{fig:pk} corresponds to $z=3$ in the simulation.  The measurement of \citet{mcdonald00} at $z=4$ in Figure \ref{fig:pk2} appears to be more consistent with \emph{slightly} cooler temperatures than their $z= 3$ measurement.  For reference, the $x_{\rm HeIII} = 0.4$ output is at $z=4$ in the reionization simulation. 

There are several systematics that may bias these interpretations.  The observations are biased in the direction of having extra power by instrumental noise and especially by metal lines, and the simulations are also biased in the direction of extra power as previously described because pressure smoothing is not properly captured and the simulation gas temperatures (prior to our post-processing) are on the low side.  \citet{lidz09} estimated that pressure smoothing could alter the power at $k\sim 0.1~$s~km$^{-1}$ by a maximum of $20\%$ (see their Fig. 29), but its impact should be significantly smaller for the \HeII\ reionization scenarios considered here where the heating has been more recent than in this case in \citet{lidz09}.  %In addition, the studies in \citet{lidz09} were not conclusive as to whether the $2\times1024^3$, $25~$Mpc/h simulation used in these calculations was in fact converged at $z=4$.  If it were not, the simulation would underpredict the total power.  
Uncertainties in the mean flux are also important for the comparison at $z=4$ \citep{lidz09}.

In conclusion, measurements of the small-scale Ly$\alpha$ forest power spectrum reveal that $15 < T_0 < 25~$kiloK at $z=3$, with tentative evidence for an decrease in temperature from $z=3$ to $4$ in the \citet{mcdonald00} measurement.  However, finer determinations would be challenging because the systematics are at the level of the model differences.  We also find that it would be difficult to detect physically motivated models for temperature fluctuations with this statistic.  

Appendix A considers the impact of the temperature models considered here on wavelet statistics.  Wavelets are also a measure of the small-scale power in the forest, and are better suited than the small-scale behavior of $P_F^{\rm los}$ for detecting temperature inhomogeneities.  Despite this advantage, Appendix A reaches similar conclusions to this section.

\subsection{Flux Probability Distribution}
\label{ss:FPDF}

\begin{figure}
\rotatebox{-90}{\epsfig{file=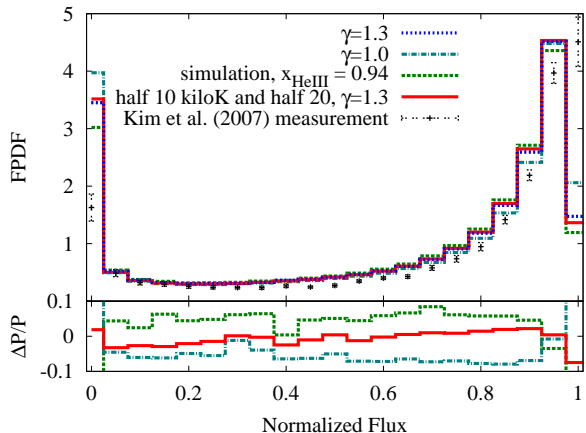, height=8.5cm}}
\caption{Top panel:  The PDF of the normalized flux. The blue dotted curve represents a model with $\gamma = 1.3$, and the teal dot-dashed curve 
for $\gamma = 1.0$.  The red solid curve represents a model with $\gamma = 1.3$, but where half the IGM has $T_0 = 10~$kiloK and the other half $T_0= 20~$kiloK.  The green dashed curve is this PDF for the temperature field from the \HeII\ reionization simulation.  The points with errorbars are the measurement of \citet{kim07}.  Bottom Panel:  The relative difference between the flux PDF in these temperature models and in the $\gamma =1.3$ model. \label{fig:PDF}}
\end{figure}

Another statistic that is often applied to the Ly$\alpha$ forest is the probability distribution function of the normalized flux (FPDF).  A distribution of temperatures at a given $\Delta_b$ will broaden the FPDF relative to the case of a single $T-\Delta_b$ relation.  In fact, measurements of the FPDF have claimed to find too many hot voids at $2 < z < 3$ compared to cosmological hydrodynamic simulations with standard, homogeneous thermal histories.  This led \citet{bolton08} and \citet{2009MNRAS.tmpL.290V} to argue that an inverted $T-\Delta_b$ relation ($\gamma - 1 < 0$) was needed to reconcile this discrepancy.  \HeII\ reionization by the observed population of quasars cannot generate an inverted relation \citep{mcquinn09}.  However, perhaps an inverted relation is not required to explain the data, but may
rather be accounted for by
additional dispersion in the temperature as conjectured in \citet{mcquinn09}.  Dispersion would have made some of the voids hotter and, as a result, more transparent to Ly$\alpha$ photons. 

The top panel in Figure \ref{fig:PDF} plots the FPDF at $z=3$.  The points with errorbars are the measured FPDF from \citet{kim07}.  The curves are this statistic calculated from $1000$ skewers through the $25/h~$Mpc simulation.  The red solid curve represents a toy model in which $\gamma = 1.3$ and where half the IGM has $T_0 = 10~$kiloK and the other half $T_0= 20~$kiloK.  The green dashed curve was calculated using the temperature at the end of the \HeII\ reionization simulation (where $x_{\rm HeIII} = 0.94$), and corresponds to the simulation snapshot in which the dispersion in temperature is approximately maximized.   The blue dotted curve represents a model with $\gamma = 1.3$ and with $T_0$ equal to the average in the $x_{\rm HeIII} = 0.94$ snapshot.  However, the dependence of this curve on $T_0$ is weak.  The bottom panel in Figure \ref{fig:PDF} plots the relative difference between the $\gamma =1.3$ model and the other temperature models featured in the top panel. 

We next estimate the magnitude by which temperature fluctuations alter the FPDF.  To proceed, we write the FPDF, ${\cal P}$, as a Taylor expansion in $\delta_{Tp7}$:
\begin{eqnarray}
{\cal P}(F) &\approx&\tilde{{\cal P}}(F) + A \, \langle \delta_{Tp7} \rangle + B \, \langle \delta_{Tp7}^2 \rangle + ...,\\
%A &\equiv& - \frac{d\tilde{\cal P}}{dF} \, \tilde{F} \, \tau, ~~ B\equiv \frac{d^2 \tilde{\cal P}}{dF^2} \, \tau^2 \, \tilde{F}^2 - \frac{d\tilde{\cal P}}{dF} \; \tilde{F} \, \left(2\,\tau - \tau^2 \right), \nonumber \\
A &\equiv&  \frac{d\tilde{\cal P}}{dF} \, \tilde{F} \, \tau - {\cal P}\left(1 - \tau \right), \nonumber \\
B &\equiv& \frac{d^2 \tilde{\cal P}}{dF^2} \, \tau^2 \, \tilde{F}^2 - \frac{d\tilde{\cal P}}{dF} \; \tilde{F} \, \left(4\,\tau - 3\,\tau^2 \right) + {\cal P} \left(2 - \tau \, \left[4 -\tau \right] \right), \nonumber 
\end{eqnarray}
where $\tilde{\cal P}$ is the FPDF without temperature fluctuations, $\delta_{Tp7}$ is the fluctuation in $T^{-0.7}$ around the average $T$-$\Delta_b$ relation, $\tilde{F}$ is flux for the average $T$-$\Delta_b$ relation, $\langle...\rangle$ represents an average at $\tilde{F}$ (roughly corresponding to fixed $\Delta_b$), and $\tau \equiv  -\log(\tilde{F})$.  Both $A$ and $B$ are of the order of ${\cal P}(F)$, and we find $A \approx 0.3$ and $B \approx -0.1$ at $F = 0.7$.  The leading order contribution from temperature fluctuations \emph{about the mean $T-\Delta_b$ relation} is $B \, \langle \delta_{Tp7}^2 \rangle$.  Plugging in $\langle \delta_{Tp7}^2 \rangle \sim 0.2^2$, which is characteristic of the simulation temperature model at $F \approx 0.7$, we estimate that temperature fluctuations should alter ${\cal P}(F)$ at the several percent level.

 The size of the residuals in the bottom panel of Figure \ref{fig:PDF} qualitatively agree with the above estimate that temperature fluctuations should produce a percent-level change in ${\cal P}(F)$.  The predictions are not in quantitative agreement (and sometimes has the incorrect sign) because we have not included the normalization to a single mean flux in our analytic expressions, which fixes the first moment of the FPDF.  We conclude that a different effect is required to create the $\approx 20\%$ \emph{suppression} of the FPDF that \citet{bolton08} finds at $F \approx 0.7$.  Additional dispersion in the temperature (even with a much larger amplitude than in our \HeII\ reionization simulations) is unable to explain the discrepancy between the observed and simulated FPDFs.
 
  Changing $\gamma$ can have a larger effect on the FPDF than adding dispersion to the temperature because, in this case, the leading order term is $A \, \langle \delta_{Tp7} \rangle$.  (Compare the $\gamma =1.3$ with the $\gamma =1.0$ curve in Figure \ref{fig:PDF}.)  Performing a more detailed comparison than considered here, \citet{bolton08} found that an inverted relation with $\gamma \approx 0.5$ provided the best fit to the \citet{kim07} FPDF measurement.  However, an inverted relation may not be the only explanation and would require an unknown heating mechanism.  The FPDF is especially sensitive to the accuracy of continuum fitting \citep{2011arXiv1103.2780L} and the efficacy of metal-line removal.

\subsection{Large-Scale Correlations}
\label{ss:LSC}

The remainder of this paper discusses large-scale correlations in the Ly$\alpha$ forest.  At scales where $\delta_F \equiv F/\bar{F} - 1$ and $\delta_T$ are small, the $3$D power spectrum of $\delta_F $ can be expressed as bias factors times the power spectra of the different sources of fluctuations.  In particular,
\begin{eqnarray}
P_{F}(\bfk) &\approx & b^2 \bigg(G^2 \, P_{\Delta}(k) + 2\,G \epsilon^{-1} \, \left[P_{\Delta \, Tp7}(k) - P_{\Delta \, J}(k) \right]  \nonumber \\
& &+  \epsilon^{-2} \, \left[P_{Tp7}(k) + P_{J}(k)\right]  \bigg),
\label{eqn:Pklg2}
\end{eqnarray}
where $\bfk$ is the Fourier wavevector, $k = | \bfk|$, $Tp7$ is shorthand for the temperature field to the $-0.7$ power, $J$ represents intensity, $P_{\Delta \, X}(k)$ is the cross power spectrum between density and the overdensity in $X$, $\mu = \hat{\bfk}\cdot \hat{n}$ where $\hat{n}$ is the line-of-sight direction, and $\epsilon \approx 2 - 0.7\,(\gamma-1)$.   The $G \equiv ( 1 + g \mu^2)$ factors arise from peculiar velocities \citep{kaiser87}, and $g$ reflects that these distortions have a different bias than density fluctuations.  Our simulations require $g \approx 1$ (Appendix~B). 

The $\epsilon$ and $\epsilon^2$ suppression of $T$- and $J$-dependent terms is approximate and results because the flux depends on $\Delta_b$, $T$, and $J$ via the combination $\Delta_F^\epsilon \,  T^{-0.7} \, J^{-1}$.  This suppression holds in linear theory, but can be violated by higher order correlations.  Equation~(\ref{eqn:Pklg2}) suggests that at scales where either $P_{Tp7}$ or $P_J$ are comparable to $P_{\Delta}$, these fluctuations can have a large effect on $P_F$.  

%Equation \ref{eqn:Pklg2} shows that it may be easier to detect temperature fluctuations at high redshift where $P_\delta$ is smaller.  %It also because $\epsilon >1$, $P_{\delta \delta_T}$ is suppressed by a smaller factor compared to $P_{\delta_T}$.  Because of the rare, stochastic nature of the bubbles during \HeII\ reionization, it is likely that  $P_{\delta \delta_T} \ll P_{\delta \delta}$.  However, during hydrogen reionization, there is likely to be a much a stronger correlation between temperature and density \citep{trac08, furlanetto09}, potentially enhancing the impact of temperature fluctuations on $P_{\delta_F}$. 

\subsubsection{Line-of-sight Correlations}
\label{ss:los}

\begin{figure}
\begin{center}
\epsfig{file=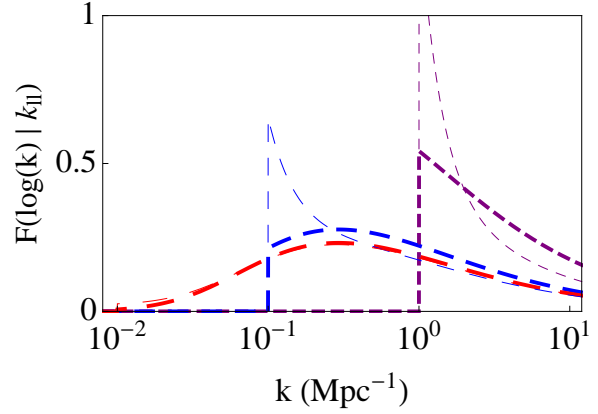, width=7.7cm}
\end{center}
\caption{Fractional contribution per $\log k$ to $P_F^{\rm los}$ from different $3$D wavevectors with modulus.  The curves (in order of decreasing dash length) are $k_{\parallel} = 10^{-2}, ~10^{-1}$, and $1$~Mpc$^{-1}$.  The thick curves are for $g = 0$ (no peculiar velocities) and the thin are for $g= 1$.  \label{fig:aliasing}}
\end{figure}

%\begin{figure}
%\begin{center}
%\epsfig{file=f9.eps, width=7.7cm}
%\end{center}
%\caption{Illustration of the impact of aliasing on line-of-sight statistics.  The solid curves are the $3$D dimensionless power spectrum of the Ly$\alpha$ flux, $k^3{P}_F/2\pi^2$, for $z=3$, $g=1$, and $\bfk_{\perp} = 0$ (black curves), of a Gaussian $3$D power spectrum with $\sigma = 20~$Mpc (blue curves), and of a scale-invariant $3$D power spectrum (red curves).  The latter two models are toy parametrizations whose spectral shapes are motivated by the temperature-fluctuation models presented earlier.  The dashed curves are the line-of-sight dimensionless power spectra for these models, $k P^{\rm los}/\pi$. \label{fig:Pkaliasing}}
%\end{figure}

\begin{figure}
\begin{center}
\rotatebox{-90}
{\epsfig{file=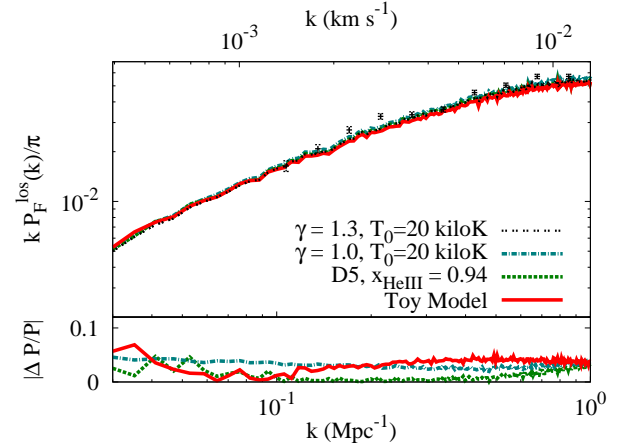, height=8.5cm}}
\end{center}
\caption{The large-scale line-of-sight dimensionless power spectrum for different temperature models at $z=2.8$ (top panel), and the relative difference between these models and the $\gamma = 1.3$ model (bottom panel).  The ``Toy Model'' curve assumes that  $\gamma = 1.3$ locally and that $30~$Mpc spheres with $T_0=20~$kiloK fill half of the volume and the rest is at $T_0=10~$kiloK.  The error bars are the SDSS measurement presented in \citet{mcdonald05b}. \label{fig:los}}
\end{figure}

Previous Ly$\alpha$ forest correlation analyses have primarily focused on the line-of-sight power spectrum and neglected correlations between sightlines.  The line-of-sight power spectrum $P_{F}^{\rm los}$ can be expressed in terms of the $3$D power spectrum $P_F$ as
\begin{equation}
P_{F}^{\rm los}(k_{\parallel}) = \int_{k_{\parallel}}^\infty \frac{dk}{2\pi} k \,P_{F} (k, k_{\parallel}/k), 
\label{eqn:Plos}
\end{equation}
where $k_{\parallel}$ is a line-of-sight wavevector.
Because $\mu \equiv k_{\parallel}/k$, the projection to $1$D suppresses the impact of large-scale peculiar velocities (and the $P_{\Delta X}$ terms; c.f. eqn. \ref{eqn:Pklg2}).  In addition, $3$D wavevectors with even $k \gg k_{\parallel}$ still contribute to $P_{F}^{\rm los}$ at $k_{\parallel}$.  Figure \ref{fig:aliasing} illustrates this aliasing effect, plotting at several $k_{\parallel}$ the 
fractional contribution per $\log k$ to $P_F^{\rm los}$ from different $3$D modes with wavevector $k$.
 The red thick dashed curve represents this for $k_{\parallel} = 10^{-1}~$Mpc$^{-1}$ and $g=0$ (the red thin dashed this for $g=1$), 
approximately the value of the \emph{3D} wavevector where $P_T$ is maximized relative to $P_\Delta$ in some of our models (Fig. \ref{fig:fluctuations}).  Thus, $P_{F}^{\rm los}(k_{\parallel})$ receives a significant contribution from approximately two decades in $k$.  This aliasing dilutes the impact of large-scale temperature fluctuations.\footnote{The calculations in Figures \ref{fig:aliasing} assume the simple form $P_F(k, k_{\parallel}) = b^2 \, [1+ g \, (k_{\parallel}/k)^2]^2 \, P_{\Delta}^{\rm lin}(k) \, \exp[-k_b T k_{\parallel}^2/m_p]$, where $P_{\Delta}^{\rm lin}$ is the linear theory overdensity power.  We take $g=1$ and $T = 20,000~$K.  We find that this functional form provides a decent approximation to the spectrum of $P_F$ in our simulations.}

   %Even though peculiar velocities buffer this aliasing effect somewhat (dashed curves in Figure \ref{fig:fluctuations}), peculiar velocities likely do not correlate strongly with temperature and, thus, act to decrease the relative contribution of temperature fluctuations to $P_{F}^{\rm los}(k_{\parallel})$ as well.

%Figure \ref{fig:Pkaliasing} provides another illustration of this effect, showing the dimensionless $3$D and $1$D power spectra (solid and dashed curves, respectively).  The black curves are the flux power without temperature fluctuations, and the red and blue power spectra are toy parametrizations characteristic of some of our temperature models.   Because $k P_F^{\rm los} \sim k$ at $k \lesssim 0.1~$Mpc$^{-1}$ owing to this aliasing, little additional cosmological information can be derived from a measurement of $P_{F}^{\rm los}$ on these scales.  In addition, this figure illustrates that the projection to $1$D is likely to hide fluctuations in the temperature in the line-of-sight power spectrum at $k \lesssim 0.1~$Mpc$^{-1}$.   At $k = 0.1~$Mpc$^{-1}$, the projection to $1$D enhances $P_{F}^{\rm los}$ by almost an order of magnitude compared to $P_{F}$ (black curves), whereas the other curves in this figure are not enhanced by this projection owing to their different shape. 

Figure \ref{fig:los} quantifies the impact of temperature fluctuations on the line-of-sight power spectrum.  The curves are $P_{F}^{\rm los}$, calculated using $22,500$ skewers of side-length $750$~Mpc/h from the $z=2.8$ snapshot of Run 1 in \citet{white10} and using different temperature models.  Each curve represents a different model for the temperature.  However, the values of $P_F^{\rm los}$ in all of the considered models differ by less than $10\%$.  %The power-spectrum of the temperature field from \HeII\ reionization differs by a maximum of $\approx 8\%$ from the curve that uses the $\gamma = 1.3$ temperature model (bottom panel).  This is a larger difference than between the $\gamma = 1.3$ model and the $\gamma = 1.0$ curve and these curves and the one for the toy temperature model, which assumes half of the IGM is at $10~$kiloK and half at $20~$kiloK.  
These small differences qualitatively agree with the results of \citet{lai06}.  

The points with errorbars in Figure \ref{fig:los} are the SDSS measurement from \citet{mcdonald05b}.  The differences between the plotted models are comparable or smaller than the  \citet{mcdonald05b} errorbars.  Thus, the impact of these temperature models almost certainly could not be detected in $P_{F}^{\rm los}$ once one marginalizes over the cosmology.  Temperature fluctuations could bias cosmological parameter determinations from $P_{F}^{\rm los}$, but at no more than the few percent-level in the considered models.   

The calculations in Figure \ref{fig:los} implicitly assume that the large-scale temperature fluctuations do not correlate strongly with the large-scale density modes due to our method for adding temperature fluctuations.  This is not always a good approximation for the calculation that uses the \HeII\ reionization simulation temperature field.  Accounting for these correlations would increase the difference with the D5 curve by approximately a factor of $2$, as can be inferred from the discussion in the Section \ref{ss:HI_3D}. 

The next section discusses the $3$D power spectrum of the flux, a statistic which avoids the aliasing issues in $P_{F}^{\rm los}$.  We show that temperature models that have almost no effect on $P_{F}^{\rm los}$ can alter the $3$D power spectrum of the flux at an order unity level.  

%For example, for $P_{\rm 3D} \sim k^{n} (1 + g \, \mu^2)^2$ and $n< -2$, one can integrate equation (\ref{eqn:Plos}), which yields
%\begin{equation}
%P_{\rm 1D}(k_{\parallel}) \sim - k_{\parallel}^{n+2} \, \left[\frac{1}{n+2} + \frac{2 \, g}{n} + \frac{g^2}{n-2} \right].
%\end{equation}
%Thus, the first term should dominate at the $0.1-1~$Mpc$^{-1}$ scales where $n \approx -2$, and the large-scale bias that results from large-scale peculiar velocities are most important where the power-spectrum is flatter (at smaller and larger scales).  We confirm that this is the case, finding that peculiar velocities alter {\bf put in numbers} In addition, at scales were $n \approx -2$, the contribution of the monopole term to $P_{\rm 1D}$ receives equal contribution per $\log k$.  {\bf Let's quote some numbers for the slope of power spectrum}  Because temperature fluctuations have a flatter spectrum, their impact on $P_{\rm 1D}$ will be suppressed compared to this for the monopole of $P_{\rm 3D}$.

\subsubsection{$3$D Correlations}
\label{ss:HI_3D}

The Sloan Digital Sky Survey III's Baryon Oscillation Spectroscopic Survey (BOSS) aims to observe the Ly$\alpha$ forest in the spectra of $1.6\times 10^5$ quasars over an area of $8000$~deg$^2$.  A future survey, BigBOSS, will push a magnitude fainter, increasing this sample by a factor of a few.\footnote{\url{http://cosmology.lbl.gov/BOSS/}, \url{http://bigboss.lbl.gov/index.html}}  BOSS's high density of quasars will allow the first measurement of the $3$D Ly$\alpha$ forest power spectrum.  A major goal of this effort is to detect the baryon acoustic oscillation (BAO) features.  The $3$D power spectrum will also be useful for studying temperature and other nonstandard sources of fluctuations in the forest because it avoids the aliasing issues that diminish the impact of large-scale correlations in $P_F^{\rm los}$.

\begin{figure}
{\epsfig{file=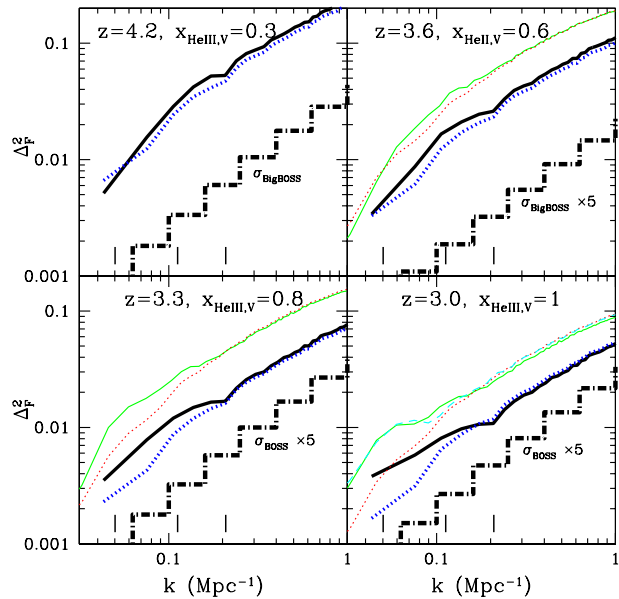, height=8.5cm}}
\caption{$3$D Ly$\alpha$ forest dimensionless power spectrum.  The dotted curves are calculated assuming $\gamma = 1.3$, and the solid curves adopt the temperature field predictions of the two \HeII\ reionization simulations.  The thick curves are from the $190~$Mpc simulation and are at the quoted redshift, and the thin curves (shown in three of the panels) are from the $430~$Mpc simulation from a snapshot that is $\Delta z \approx 0.3$ higher.  The vertical line segments mark the centers of the three BAO features that upcoming quasar surveys are targeting as a standard ruler.  The bottom panels also include a piecewise curve representing the measurement error \emph{multiplied by 5} of BOSS for the signal in each piecewise region as described in the text.  The top panels include similar sensitivity curves but for BigBOSS.  The cyan thin dashed curve in the bottom right panel also includes the prediction for $\Delta_F^2$ in a toy temperature model that has $30$~Mpc spheres at twice the temperature filling half of the volume.
\label{fig:3dspectrum}}
\end{figure}

Figure \ref{fig:3dspectrum} quantifies the effect of the temperature fluctuations in the \HeII\ reionization simulations on $P_F$.  The thick set of curves are calculated from the $190$~Mpc simulation, and the thin (shown in three of the panels) are from the $430~$Mpc one.  The different normalization of the curves in the two simulations results primarily because of the different redshifts of the snapshots.  The redshift of the $430~$Mpc simulation curves is $\Delta z \approx 0.3$ higher than the quoted redshift because \HeII\ reionization occurs slightly earlier in this simulation (owing to fewer recombinations in this simulation).  When calculating these curves, we normalize to the observed mean flux at the respective redshift.   The solid curves are $\Delta_F^2$ computed using the temperature field from the \HeII\ reionization simulation at the quoted $x_{\rm HeIII}$ and, for comparison, the dotted curves show $\Delta_F^2$ for a power-law $T-\Delta_b$ relation with $\gamma = 1.3$.  Changing $\gamma$ or $T_0$ has little effect on the latter curve.  

Figure \ref{fig:3dspectrum} illustrates that temperature fluctuations can significantly alter $\Delta_F^2$ on large scales, changing its amplitude by as much as a factor of $2$.  In addition, the relative impact of temperature fluctuations at fixed $x_{\rm HeIII}$ largely agree between the small- and large-box simulations.  This agreement is suggestive that the impact of temperature fluctuations on these scales is not affected by how well the small-scale features in the forest are resolved.  The bottom right panel shows for comparison $\Delta_F^2$ predicted for our toy model with $30$~Mpc bubbles at twice the temperature filling half of space in the larger box (cyan thin dashed curve).  Interestingly, even though the temperature fluctuations are much larger in this model than in the simulations (see Fig. \ref{fig:fluctuations}), their effect on $\Delta_F^2$ is comparable to that in the simulation temperature models.

  The BOSS quasar survey is forecast to measure $\Delta_F^2$ at $z\approx 3$ to $\sim 10\%$ accuracy in bins with $\Delta k \sim k$ in the range $10^{-2} < k < 1~$Mpc$^{-1}$, precision beyond what would be required to detect this extra power in some of our models.  The bottom panels in Figure \ref{fig:3dspectrum} includes an estimate discussed in \citet{mcquinnwhite} for the measurement error of BOSS \emph{multiplied by 5}, assuming $8000~$deg$^{-2}$ over a region of depth $500~$Mpc.  These curves are calculated assuming $\bar n_{\rm eff} = 3\times 10^{-4}~$Mpc$^{-2}$ for the $z=3$ panel and $\bar n_{\rm eff} = 2\times 10^{-4}~$Mpc$^{-2}$ for the $z=3.3$ (corresponding to $\approx 2-3~$per deg$^{2}$), where $n_{\rm eff}$ is a noise weighted number density of quasars on the sky and defined in \citet{mcquinnwhite}.  The sensitivity scales as $n_{\rm eff}^{-1}$.  These number densities take the quoted magnitude limit for BOSS of $m=22$, assuming that all quasars below this magnitude are targeted and that $S/N =1$ in $1~$\AA\ pixels at $m=22$.   With these assumptions, BOSS will be several times more sensitive to $P_F$ at $z=2-2.5$ compared to at $z=3$.  In addition, the vertical line segments in this figure mark the centers of the three BAO features that the BOSS and other upcoming quasar surveys are targeting as a standard ruler.  Temperature fluctuations have the potential to bias the measurement of this feature unless a fairly general functional shape is assumed for the continuum on which these features sit.

BigBOSS aims to obtain a denser sample of quasars, and we forecast that it will be able to constrain $\Delta_F^2$ to $z \approx 4$ if its survey strategy is similar to that of BOSS, and possibly to higher redshifts if this strategy is optimized to go deeper \citep{mcquinnwhite}.  The dot dashed curves in the top panels in Figure \ref{fig:fluctuations} show what BigBoSS can achieve, assuming that it obtains spectra for quasars that are a magnitude fainter than BOSS such that $\bar n_{\rm eff} = 5\times 10^{-4}~$Mpc$^{-2}$ at $z=2.6$ and $1\times10^{-4}~$Mpc$^{-2}$ at $z=4.2$. 

Peculiar velocities are not included in the calculation of the curves in Figure \ref{fig:3dspectrum} and the gas is assumed to be in the Hubble flow.  Peculiar velocities reduce the importance of temperature fluctuations.  However, peculiar velocities do not contribute to the power in modes transverse to the line of sight, and, thus, we expect that the differences seen in Figure \ref{fig:3dspectrum} should be representative of the effect on transverse modes and the monopole component of $\Delta_F^2$.  Observations will be most sensitive to the $\Delta_F^2$ monopole (Section \ref{sec:angular}).

\begin{figure}
{\epsfig{file=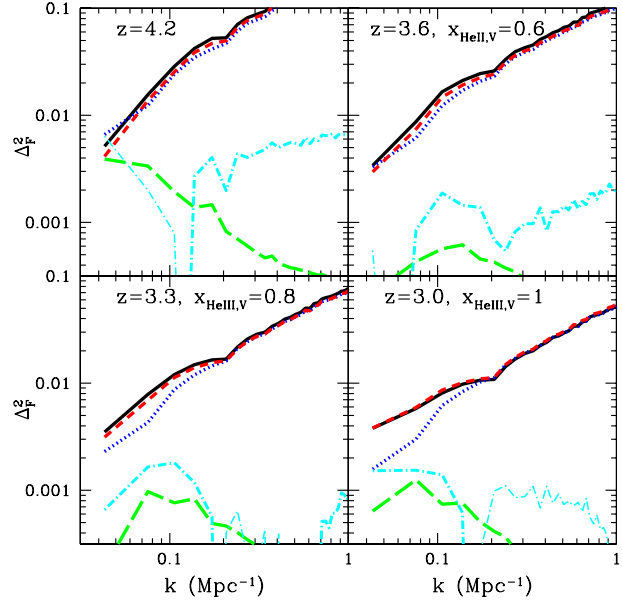, width=8.5cm}}
\caption{Comparison of the predictions for $\Delta_{F}^2$ using equation \ref{eqn:Pklg2} (red dashed curves) and the full numerical result from the $190~$Mpc box (black solid curves).  The blue dotted curves are $\Delta_{F}^2$ for a power-law $T-\Delta_b$ relation with $\gamma = 1.3$, and the green dashed curves and cyan dot-dashed curves are respectively the contribution proportional to $P_{Tp7}$ and $|P_{Tp7 \Delta}|$.  In particular, the thin cyan dot-dashed curves are $-2 \, b\, \epsilon^{-1} \, P_{Tp7 \Delta}$ and the thick are $2 \, b \, \epsilon^{-1} P_{Tp7 \Delta}$.  In each panel, the sum of the blue, green, and $\pm$ the cyan dot-dashed curve yields the red dashed curve.  
\label{fig:modelcomp}}
\end{figure}

Figure \ref{fig:modelcomp} demonstrates that equation (\ref{eqn:Pklg2}) provides a reasonable description of the impact of temperature fluctuations in the reionization simulations.  This figure compares the predictions for $\Delta_{F}^2$ from equation (\ref{eqn:Pklg2}) (red dashed curve) with that of the full numerical result (black solid curve).  The red dashed curve is constructed by adding each contribution that appears in equation (\ref{eqn:Pklg2}) separately, and the value of $b^2$ in this equation was derived by taking the ratio of $\tilde{P}_F$ to $P_\Delta$. 
The blue dotted curve is $\Delta_{F}^2$ for a power-law $T-\Delta_b$ relation with $\gamma = 1.3$, the green long dashed curve and thick cyan dot-dashed are respectively $(b/\epsilon)^2 \, \Delta_{Tp7}^2$ and $2 b^2/\epsilon \, \Delta_{Tp7 \Delta}^2$.  (The thin cyan curves are $-2 b^2/\epsilon \,  \Delta_{Tp7 \Delta}^2$.)  The field $\delta_{Tp7}$ is taken to be the overdensity in temperature around the mean $T-\Delta_b$ relation, a slightly different definition than used previously and that is motivated in Appendix~C.  The sum of the blue dotted, green long dashed, and $\pm$ the cyan dot-dashed curve yields the red dashed curve, which is remarkably similar to the black solid curve.  (See Appendix~C for a complementary analytic description of the impact of temperature fluctuations.)

The history of $P_{Tp7 \Delta}$ in the \HeII\ reionization simulation is complicated.  This function is less than zero at the beginning of \HeII\ reionization and at the largest-scales  because quasars are heating their immediate surroundings (top left panel, Fig.  \ref{fig:modelcomp}).  Later, it becomes positive as the large-scale void regions are ionized.  Constraining this nontrivial evolution from observations of $\Delta_F^2$ would inform \HeII\ reionization models.  
%Similarly, the scale dependence of $P_{Tp7}$ depends largely on the quasar luminosity function and quasar lifetimes.

The contribution from $P_{Tp7 \Delta}$ to $\Delta_F^2$ is of comparable import to the contribution from $P_{Tp7}$ in the four cases considered in Figure \ref{fig:modelcomp}, despite $P_{Tp7 \Delta}$ being smaller than $P_{Tp7}$.  This occurs because $P_{Tp7 \Delta}$ is enhanced relative to $P_{Tp7}$ by a factor of $2 \, \epsilon$ in equation (\ref{eqn:Pklg2}).  This enhancement explains why the temperature fluctuations in the simulations produce a comparable change in $\Delta_F^2$ to that in our toy temperature model, despite the toy model having a much larger $P_T$.  If the \HeII\ ionizers were more abundant than assumed here (as would be the case if the faint end of the quasar luminosity function were steeper; \citealt{furlanetto07b}), $P_{Tp7 \Delta}$ could be significantly larger, increasing the impact of temperature fluctuations on $P_F$.    % Models where temperature fluctuations are a few times smaller than in our simulation could also likely be detected with BOSS, especially if these fluctuations were present at the redshift sweet-spot for BOSS, $z\approx 2.5$.
%Similarly, it is likely that hydrogen reionization resulted in a larger $P_{Tp7 \Delta}$ \citep{furlanetto09, cen09} such that the impact of the temperature fluctuations it imparted was even larger in $\Delta_F^2$.  

In conclusion, temperature fluctuations could have a significant effect on the $3$D power spectrum in the forest and would be detectable in future $3$D Ly$\alpha$ forest analyses such as with BOSS.  Our simple analytic description for their impact is extremely successful.  

\section{Intensity Fluctuations}
\label{app:Gfluc}
\label{sec:Gamma}

Several studies have investigated the impact of intensity ($J$) fluctuations on the Ly$\alpha$ forest \citep{zuo92, zuo92b, meiksin04, croft04, mcdonald05, furlanettoJfluc, 2009MNRAS.400.1461M}.   Here we attempt to understand how intensity fluctuations could affect $3$D correlations in the forest.  \citet{meiksin04} and \citet{croft04} concluded that $J$ fluctuations have a small effect at $10-100$ km~s$^{-1}$ separations in the Ly$\alpha$ forest correlation function, but found that they could change this statistic at the $10$s of percent-level on larger scales.  Interestingly, \citet{white10} found that intensity fluctuations could have an order-unity impact on $3$D correlations.  A detection of intensity fluctuations would constrain the rarity of the sources of ionizing photons, and it would constrain the contribution of quasars versus that of galaxies to the metagalactic ionizing background.

Intensity fluctuations (equivalent to $\Gamma_{-12}$ fluctuations) alter the $3$D Ly$\alpha$ forest power spectrum on linear scales as
\begin{equation}
P_F(\bfk) \approx \tilde{P}_F(\bfk) + b^2 \left[\epsilon^{-2} \, P_{J}(k) - 2 \, \epsilon^{-1}  (1+g\mu^2)\, P_{\Delta J}(k) \right],
\label{eqn:PFJ}
\end{equation}
(c.f. eqn. \ref{eqn:Pklg2}) where $\tilde{P}_F(\bfk)$ is the flux power-spectrum without $J$ fluctuations.  If we assume a Euclidean space in which all \HI\ ionizing photons experience a single attenuation length $\lambda$ and treat the quasars as continuously shining lightbulbs, then 
\begin{equation}
P_J =   \left( \frac{\arctan(\lambda k)}{\lambda k} \right)^2 \left(\frac{\langle L^2 \rangle}{\bar{n} \, \langle L \rangle^2 } +  b_q^2 \, P_{\Delta}(k) \right),   \label{eqn:PJ}
\end{equation}
(which is similar to the expression for $P_J$ in \citealt{2009MNRAS.400.1461M}) and
\begin{equation}
P_{\Delta  J}  =  \left( \frac{\arctan(\lambda k)}{\lambda k} \right)\, b_q\, P_{\Delta}(k).
  \label{eqn:PDJ}
\end{equation}
Here, $P_J$, is the power spectrum of intensity fluctuations, $P_{\Delta J}$ is the cross correlation between intensity and density, $b_q$ and $\bar{n}$ are respectively the luminosity-weighted bias and $3$D number density of the sources, and $(\lambda k)^{-1}\arctan(\lambda k)$ is (up to a constant factor) the Fourier transform of $r^{-2} \exp[-r/\lambda]$.\footnote{
 A commonly used formula for the correlation function of intensity fluctuations, $\xi_{J}$, is
\begin{equation}
\xi_J(r) = \frac{1}{3 \, N} \;  \frac{\langle L^2 \rangle}{\langle L \rangle^2}  \; \frac{\lambda}{r} ~ I_J(\frac{r}{\lambda}),
\end{equation}
with
\begin{equation}
I_J(u) = 2 \int_0^\infty dv \, \frac{v }{\sinh v} \, \exp \left[ - u \, \frac{1 + e^{-v}}{1- e^{-v}} \right],
\end{equation} 
and $N = 4 \pi \lambda^3 \bar{n}/3$.  Note that this formula for $\xi_J$ (derived by fairly complicated means in \citealt{zuo92b}) is just the convolution of $r^{-2} \,\exp[-r/\lambda]$ with itself (times a factor that depends on the luminosity function).  Put another way, it is the Fourier transform of the Poissonian term in equation (\ref{eqn:PJ}).  %Previous studies found it to be of note that the radial dependence of $\xi_{J}$ does not depend on the luminosity function, a result that can be understood trivially from equation (\ref{eqn:PJ}).
}  

  \begin{figure}
  {\epsfig{file=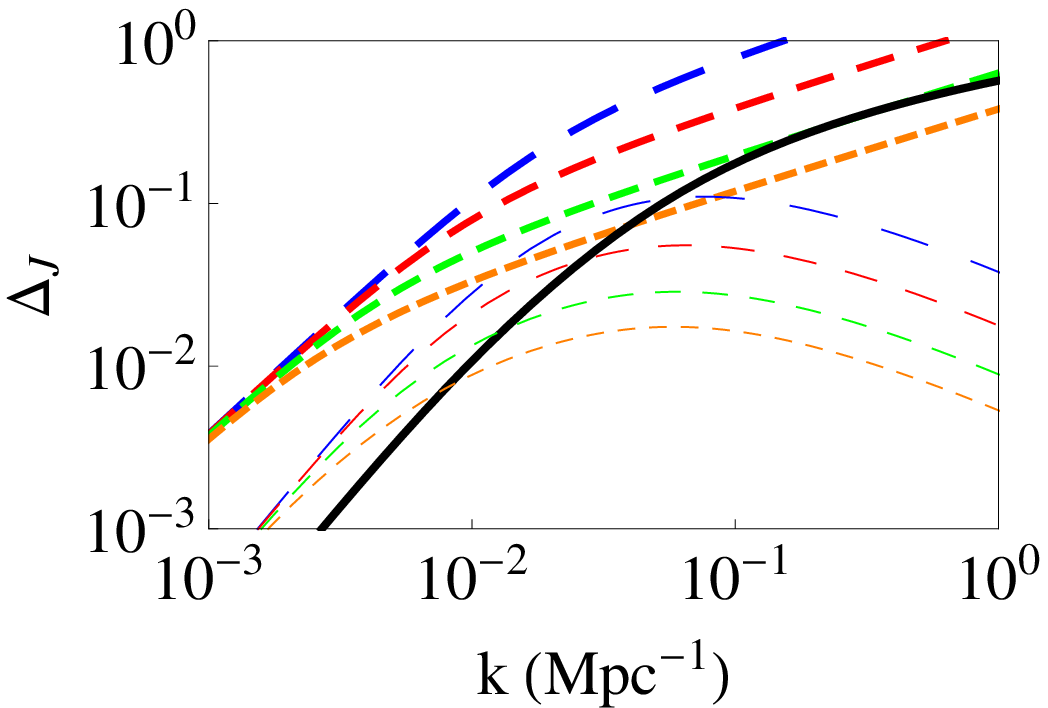, width=8cm}}
    {\epsfig{file=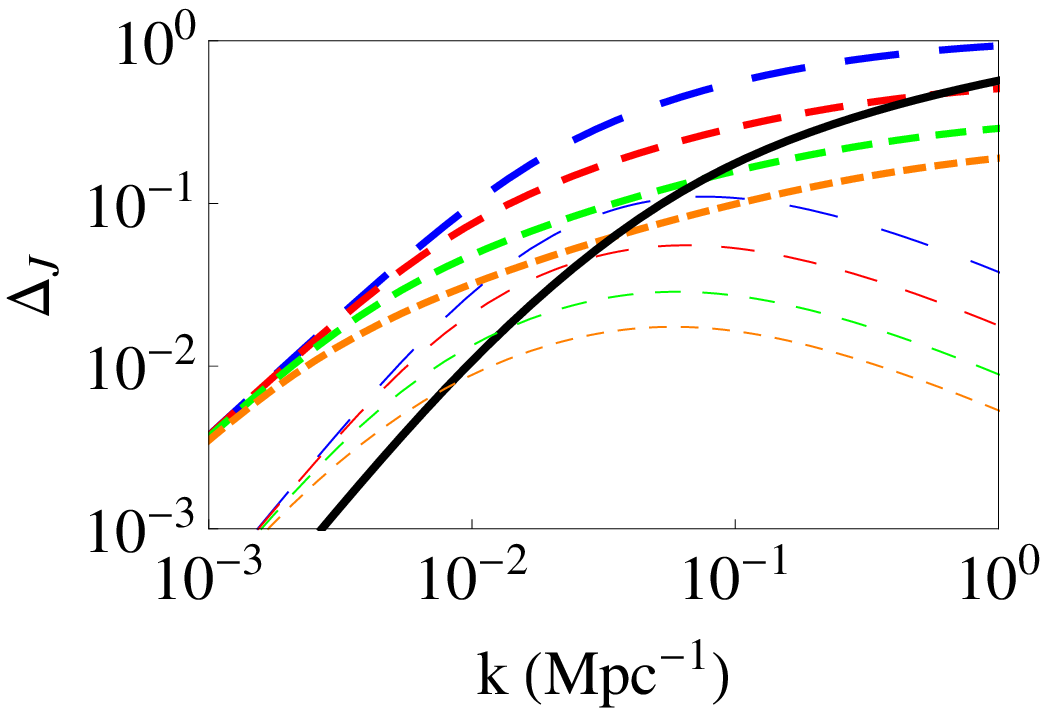, width=8cm}}
  \caption{Top panel:  Dimensionless power spectrum of intensity fluctuations to the $1/2$ power.  The thick dashed curves are the Poisson component, and the thin dashed are the clustered component.  Both sets of curves assume $\lambda = 500, 300, 150,$ and $70~$Mpc (from bottom to top), corresponding roughly to the mean free path at $z=2, 3, 4,$ and $5$.  For reference, the thick black curve is $[k^3 \, P_\Delta/2 \pi^2]^{1/2}$ in linear theory at $z=3$.
  The Poisson-component curves assume $\bar{n} = 10^{-4}~$Mpc$^{-3}$ and $\langle L^2 \rangle/\langle L \rangle^2 \approx 30$, which set these curves' normalizations.  The clustering curves assume $b_q = 3$ and $z=3$.  Bottom panel:  Same as the top panel but where the power from the proximity region is smoothed out for the Poissonian curves as described in the text. Note that $10^{-2}~$Mpc$^{-1}$ is approximately the smallest wavevector that can be estimated from the forest.
   \label{fig:Jfluc}}
  \end{figure}

Finite quasar lifetimes alter $P_J$ when the lifetime $t_{\rm q}$ is comparable to or shorter than $\lambda/c$, or $300$~Myr for $\lambda = 100$~proper Mpc.  The variance of $J$ does not depend on $t_{\rm q}$ when normalizing to a single luminosity function, but finite lifetimes can reduce large-scale $J$ correlations because $J$ becomes uncorrelated in regions separated by distances greater than $>c \, t_{\rm q}$.  These effects can be included in equation (\ref{eqn:PFJ}) by substituting a more complicated window function for $(\lambda k)^{-1}\arctan(\lambda k)$.  However, the primary effect is for $P_J$ to become white at wavelengths greater than $c \, t_{\rm q}$ rather than at $\sim\lambda$.  Thus, detecting intensity fluctuations on $100~$Mpc scales would place interesting constraints on quasar lifetimes.  In addition, light travel effects alter the isotropy of the $J$ fluctuations because the emission will have travelled further in the line-of-sight direction.  This would result in additional $\mu$-dependent terms in equation (\ref{eqn:PJ}).

The top panel in Figure \ref{fig:Jfluc} plots the Poisson and clustering components of $\Delta_J \equiv [k^3 \, P_J/2\pi^2]$ (the first and second terms in eqn. \ref{eqn:PJ}) for infinite lifetimes.  The thick dashed curves are the Poisson component for $\lambda = 500, 300, 150,$ and $70~$Mpc (in order of decreasing amplitude).   For reference, these $\lambda$ correspond roughly to its measured value at $z=2, 3,4$ and $5$ \citep{faucher08, prochaska09}.  Our calculations assume that the quasar luminosity function has the form $\Phi(L) \sim ([L/L_*]^{\alpha} + [L/L_*]^{\beta})^{-1}$ with a cutoff $3$ decades above and below $L_*$, where $\alpha = 1.5$ and $\beta = 3.5$.  These choices result in $\langle L^2 \rangle/\langle L \rangle^2 \approx 30$.  Changing both the upper and lower cutoff by a factor of $10$ and in opposite directions results in a factor of $3$ change in $\langle L^2 \rangle/\langle L \rangle^2$.   Furthermore, the curves in the top panel in Figure \ref{fig:Jfluc} use $\bar{n} = 10^{-4}~$Mpc$^{-3}$, representative of quasars at $z = 2-3$.  The \citet{hopkins07a} quasar luminosity function yields $\bar{n}= \{1.2\times 10^{-4},~ 9\times 10^{-5},~5\times 10^{-5}, ~2\times 10^{-5} \}$~Mpc$^{-3}$ at $z = \{2.5, ~3, ~4, ~5\}$.  
 
For three of the four cases featured in the top panel in Figure \ref{fig:Jfluc}, the power in the Poisson component is \emph{always} larger than the power in the density field (the black solid curve).  However, the normalization of the Poissonian component of $\Delta_J$ is highly uncertain.   The Poisson curves in the top panel of Figure \ref{fig:Jfluc} are roughly a factor of $3$ below the corresponding estimate in \citet{furlanettoJfluc} in part because \citet{furlanettoJfluc} used $\beta = 2.9$ for which $\langle L^2 \rangle/\langle L \rangle^2 \approx 100$ with our luminosity cutoffs.  Most investigations of $J$ fluctuations on the forest have used values for $\beta$ that are more similar to the value used here \citep{meiksin04, croft04, mcdonald05}.  However, \citet{furlanettoJfluc} argued that $\beta = 2.9$ provides a better fit to the \citet{hopkins07a} luminosity function.
 
%In addition, $\bar{n}$ decreases above and below $z=2.5$ if quasars are the sources.  The \citet{hopkins06} quasar luminosity function yields $\{1.2\times 10^{-4},~ 9\times 10^{-5},~5\times 10^{-5}, ~2\times 10^{-5} \}$~Mpc$^{-3}$ at $z = \{2.5, ~3, ~4, ~5\}$.  However, if galaxies dominate the ionizing emissivity (as several studies have argued was the case at $z\gtrsim 4$; e.g. \citealt{faucher08}), the amplitude of Poisson term will be reduced by the fraction of ionizing photons from quasars.
 
The thin dashed curves in the top panel in Figure \ref{fig:Jfluc} represent the contribution to $\Delta_J$ from source clustering at $z=3$ for $b_q=3$ and for the same $\lambda$ as the corresponding thick curve.  These curves fall below the Poisson component with the same $\lambda$.  However, the clustering component of intensity fluctuations will likely be the dominant source of $J$ fluctuations if galaxies are the source of ionizing photons (as several studies have argued might be the case at $z\gtrsim 4$; e.g. \citealt{faucher08}).  Because the clustered component also affects the flux field via $P_{\Delta J}$, this will enhance its effect beyond its contribution to $P_J$ and also allow this contribution to be separated via its distinct angular dependence (Section \ref{sec:angular}).  In addition, the power in the clustering component of $J$ is always larger than that in the density --- the standard source of fluctuations in the forest --- at $k \lesssim \lambda^{-1}$ since $b_q > 1$.  These scales become observable in the Ly$\alpha$ forest at $z \gtrsim 3$.

\subsection{Effect of Intensity Fluctuations on $P_F$}

  \begin{figure}
  \rotatebox{-90}{\epsfig{file=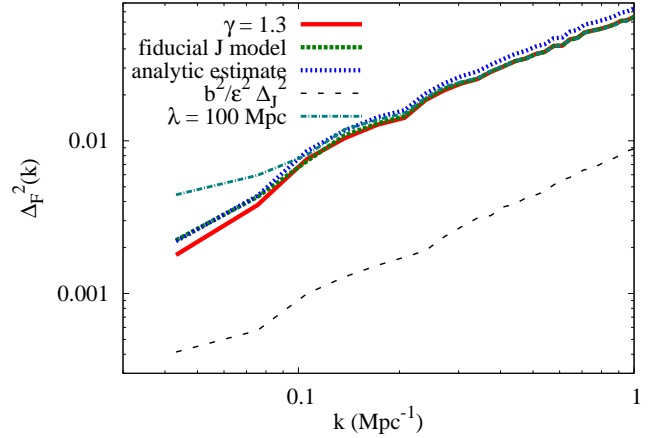, height=8.8cm}}
  \caption{Impact of intensity fluctuations on $\Delta_F^2$ at $z=3$, calculated using the $190~$Mpc simulation.  The solid red curve represents this for a uniform intensity field, and the blue dotted represents this for a fluctuating field with $\lambda  = 300~$Mpc, $\langle L^2 \rangle/ \langle L \rangle^2 = 30$, and $\bar{n} = 10^{-4}~$Mpc$^{-3}$.  The green dashed curves are the predictions by adding $\Delta_J^2$ to $\tilde{\Delta}_F^2$ as specified by equation (\ref{eqn:PFJ}).  The dashed black curve is $b^2/\epsilon^2 \Delta_J^2$ in this model, and the teal dot-dashed curve is the same as the green dashed but for a model with $\lambda = 100~$Mpc.    \label{fig:PFJ}}
  \end{figure}

The previous section showed that $P_J$ can be comparable to $P_\Delta$ in the Ly$\alpha$ forest.  If our simple analytic expression is correct (eqn. \ref{eqn:PFJ}), this implies that $J$ fluctuations significantly increase $P_F$ beyond $\tilde{P}_F$.  However, previous numerical studies of the impact of intensity fluctuations on $P_F^{\rm los}$ have considered models where the power in $J$ was comparable to or larger than the power in $\Delta_b$, and these studies found that these fluctuations had a small effect on $P_F^{\rm los}$ for $k > 0.1~$Mpc$^{-1}$  (e.g., \citealt{croft04, mcdonald05}).  Here we try try to resolve this apparent discrepancy.

Figure \ref{fig:PFJ} features numerical calculations of the impact of intensity fluctuations on $\Delta_F^2$.  The thin black dashed curve is the contribution to the total power from intensity fluctuations using equation (\ref{eqn:PFJ}) and the fiducial model for $J$ fluctuations with $\lambda = 300~$Mpc.  To generate the $J$ field, we have convolved a random distribution of quasars which have the said luminosity function with the function $r^{-2}\, \exp[-r/\lambda]$.  This ignores the clustering contribution, which Figure \ref{fig:Jfluc} suggests is smaller but non-negligible in this model.   The solid red curve is $\Delta_F^2$ without intensity fluctuations, and the dashed green is the fully numerical calculation of the effect of intensity fluctuations in this model.  The blue dotted curve is the prediction of equation (\ref{eqn:PFJ}) -- the addition of the dashed black and solid red curves.  The analytic model does poorly at capturing the impact of intensity fluctuations at $k > 0.1~$Mpc$^{-1}$, but does better at smaller wavevectors.  The teal dot-dashed curve is the same as the green dashed but for a model with $\lambda = 100~$Mpc such that the intensity fluctuations are larger and thus have a larger impact on $\Delta_F^2$.

The poor performance of the analytic model at $k > 0.1~$Mpc$^{-1}$ owes to the diverging character of $J$ in quasar proximity regions.   Within the proximity region of a quasar (within the distance $r_p(L) \equiv [4 \pi \langle L \rangle/L \,\bar{n} \, \lambda]^{-1/2}$), the effect on $\Delta_F^2$ will be suppressed beyond the prediction in equation (\ref{eqn:PFJ}) by the exponential transformation from $\tau_{\rm Ly\alpha}$ to transmission.  Because $\tau_{\rm Ly\alpha} \sim (1 + [r_p/r]^2)^{-1}$, the small-scale transmission power from proximity regions is damped exponentially with a kernel similar to that of a Gaussian with s.d. $\approx r_p^{-1}/\sqrt{2}$.  The bottom panel in Figure \ref{fig:Jfluc} is the same as the top panel, but where for the Poisson term in $P_J$ given by $k^{3/2} \arctan( \lambda k)/\lambda k$ has been forced to transition to a constant function of $k$ with the transition occurring at $k = r_p(L)^{-1}/\sqrt{2}$.  This operation is meant to approximate the convolution of $\arctan( \lambda k)/\lambda k$ with the Gaussian-like shape of the proximity region.  The power in $J$ is significantly affected by this operation at the largest $k$ that are shown.  The $k$ where the proximity-region damping occurs are roughly $k \gtrsim r_p(\langle L^2 \rangle^{1/2})^{-1}$, which correspond to $k \gtrsim 0.1~$Mpc$^{-1}$ in our fiducial model.  Smaller $k$ correspond to where the analytic model provides a good description in Figure \ref{fig:PFJ}.  %We have also verified this convergence in a model with a several times larger $P_J$ than our fiducial model such that their impact on $P_F$ is several times larger. (Although, $r_p$ is larger in this model such that the analytic and numerical curves converge at even smaller $k$.)  

%A measurement of excess power owing to $J$ in $P_F$ at $k < r_p(\langle L^2 \rangle^{1/2})^{-1}$ would constrain the quantity $\langle L^2 \rangle /[\langle L \rangle^2  \,\lambda^2  \, \bar{n}]$, and the location of the knee in $\Delta_J$ constrains the value of ${\it min}[\lambda, c t_{\rm q}]$.  

Density fluctuations become larger than those in $J$ at $k > 0.3~$Mpc$^{-1}$ in three of the Poissonian models in the bottom panel of Figure \ref{fig:Jfluc}  (and the intensity power is also suppressed by an additional factor of $\epsilon^2$ in $\Delta_F^2$).  Wavevectors with $k > 0.3~$Mpc$^{-1}$ correspond roughly to those where $P_F^{\rm los}$ has been measured.  As with temperature fluctuations, the impact of intensity fluctuations are further diluted by aliasing in $P_F^{\rm los}$ such that $3$D fluctuations are better suited for detecting them.  Thus, it is unlikely that the intensity fluctuations in three of these models would have produced a detectable imprint on $P_F^{\rm los}$.  In fact, we have calculated the impact of our fiducial $J$ model on $P_F^{\rm los}$ in the $25/h~$Mpc box and, because of this small-scale damping, found no appreciable effect owing to these damping and aliasing effects.

At $z\gtrsim4$, it is conceivable that intensity fluctuations contribute a substantial portion of the power even in $P_F^{\rm los}$ because both $\lambda$ and $\bar{n}$ are smaller.  Interestingly, \citet{mcdonald05b} constrained the $z=4$ mean flux from power spectrum measurements to be $10-20\%$ higher than the more direct measurement in \citet{faucher07}.  Such a disparity could potentially occur if an unaccounted source of fluctuations were contributing to $P_F^{\rm los}$.  Some of the intensity models considered in \citet{mcdonald05b} [their Fig. 5] and more generally in models where $\langle L^2 \rangle /[\langle L \rangle^2 \, \lambda^2 \, \bar{n}] \gtrsim 10^2~$Mpc are able to produce a tens of percent change in the power at all scales.

There are several possibilities for how the impact of intensity fluctuations on $P_F$ can be distinguished from the fluctuations in the temperature.  Temperature fluctuations should peak in amplitude at $z=2.5-3$ if \HeII\ reionization were ending around these redshifts.  In contrast, intensity fluctuations are likely to have increased monotonically with redshift.  In addition, the amplitude of $P_J$ is set by the factor $\langle L^2 \rangle /[\langle L \rangle^2  \,\lambda^2  \, \bar{n} _{\rm 3D}]$, which can potentially be constrained by ancillary observations. 

\section{Separating the Components of $P_F$}
\label{sec:angular}

The signal we are interested in has the form $P_F = P_{0} + 2 g \, \mu^2 P_{1} + g^2 \mu^4 P_{2}$ in linear theory, where the $P_i$ are functions of only $k \equiv | \mathbf{k} |$ and are given by $P_{0} = b^2 \,P_{\Delta} + b^2 \,\epsilon^{-2} \, [P_{J} + P_{Tp7}]$, $P_1 = 2 \, b^2\, P_{\Delta} + 2 \,b^2 \,\epsilon^{-1} \, [P_{\Delta J} + P_{\Delta Tp7}]$, and $P_2 = b^2 \,P_{\Delta}$.  Thus, temperature and intensity fluctuations affect $P_{0}$ and, to the extent they correlate with density, $P_{1}$.  (Light-travel effects can generate additional $\mu^{2n}$ terms that we do not consider.)  Here we investigate the sensitivity of a survey to the terms in this angular decomposition.  Detecting these angular components may prove easier in the Ly$\alpha$ forest than it has for galaxy surveys because the velocity bias $g$ is likely to be larger in the Ly$\alpha$ forest than in galaxy surveys (for which $g = 1/b_g$, where $b_g$ is the galaxy bias).

 The Fisher matrix for the parameters $P_{i}$ is
\begin{equation}
\matF_{\rm i j} = \sum_{\rm k-shell}  \frac{1}{{\rm var}[P_F(\bfk)]} \, \frac{dP_F(\bfk)}{dP_i}  \frac{dP_F(\bfk)}{dP_j}, 
\end{equation}
where the sum is over ${\cal N}_{k}$ independent elements in a shell in $k$-space centered at distance $|\bfk|$ from the origin.  In addition, ${\rm var}[P_F(\bfk)]$ is the variance on an estimate of $P_F$ in a pixel centered on $\bfk$.  

Let us first approximate ${\rm var} [P_F(\bfk)]$ as isotropic, which is the relevant limit for BOSS and, more generally, for all surveys with less than $100$ Ly$\alpha$ forest spectra per degree \citep{white10, mcquinnwhite}.  In this limit, the error on a measurement of $P_i$ is
\begin{equation}
\boldsymbol{\sigma}(k) \equiv {[\matF^{-1}]^{1/2}}_{ii}  = \frac{{\rm  var} [P_F(k)]^{1/2}}{\sqrt{{\cal N}_{k}}} \; (1.5, \; 4.0 \, g^{-1}, \;10.7 \, g^{-2}).
\end{equation}
Also informative is the cosmic variance limit in which ${\rm  var} [P_F(\bfk)] = 2\, b^4 P_\Delta(k)^2 \; (1+ g\mu^2)^4$, and for $g=1$ results in
\begin{equation}
\boldsymbol{\sigma}(k) = \sqrt{\frac{2}{{\cal N}_{k}}}\, b^2 \, P_\Delta (k) \; (2.1, \;20.4, 28.2).
\end{equation}
In both limits, a survey detects $P_0$ at the highest significance.  $P_1$ reveals how the temperature and photoionization fluctuations correlate with those of density.  Because BOSS can constrain $P_F$ in bins of $\Delta k \sim k$ at $z\sim 2.5$ at the $1\%$ level, it should also be able to place $10\%$-level constraints on $P_1$.

%\subsubsection{Sensitivity of Different Surveys}

%To detect temperature or intensity fluctuations, a survey must be able to at the very least detect order unity changes in $P_F$.  A noiseless survey is cosmic variance-limited for modes smaller than $\bar{n}_{\rm 2D}^{-1} P_{F}^{\rm los}(k_{\parallel}) < P_F(k_{\parallel}, \bfk_\perp)$, which roughly equates to $k < 2 \pi \, \bar{n}_{\rm 2D}^{1/2}$.  To be cosmic variance-limited at $k= 0.1~$Mpc$^{-1}$ (where we find temperature fluctuations are most important) requires $\bar{n}_{\rm 2D} \gtrsim 0.1~$Mpc$^{-2}$.  The ability to measure modes with $k > 2 \pi \, \bar{n}_{\rm 2D}^{1/2}$ falls off rapidly because $P_{F, \rm los}/P_F \sim k^{-2}$ (even neglecting the additional covariances that arise when $P_{F}^{\rm los}$ becomes important). 

\section{Conclusions}

Significant temperature fluctuations with $\Delta T/T \approx 1$ were
likely imprinted in the high-redshift IGM by reionization processes,
and they would have lasted for roughly a Hubble time thereafter.  Temperature fluctuations would have been imprinted at $z\sim 3$ if
\HeII\ reionization were ending near this redshift and, thereby,
affected the majority of Ly$\alpha$ forest spectra.  In addition, if
hydrogen reionization ended at $z\approx 6$, remnant temperature
fluctuations could still be observable in the forest from redshifts as low as $z =4$
(e.g., \citealt{cen09}).  This paper investigated the detectability of
these fluctuations.
%, with a focus on the temperature history predicted from the \HeII\ reionization simulations of \citet{mcquinn09}.

Previous studies have shown that the $z=2-4$ IGM cannot have been half filled with $\sim 10~$Mpc patches of temperature $20~$kiloK and the rest at $10$~kiloK \citep{2002MNRAS.332..367T, lidz09}.  However, we showed that half with $T_0 = 25~$kiloK and half with $T_0 = 15~$kiloK is consistent with recent Ly$\alpha$ forest analyzes.  Such a temperature distribution is also close to what is found near the end of \HeII\ reionization in the radiative transfer simulations of \citet{mcquinn09}.  If fact, we showed that the temperature history in the fiducial simulation of \HeII\ reionization in \citet{mcquinn09} is grossly consistent with previous Ly$\alpha$ forest measurements, although the observations tentatively favor less evolution in the mean temperature.  The temperature fluctuations in this simulation do not alter the small-scale Ly$\alpha$ forest power spectrum (Section \ref{ss:sslos}), the PDF of the normalized flux (Section \ref{ss:FPDF}), the large-scale line-of-sight power spectrum (Section \ref{ss:LSC}), the wavelet PDF (Appendix A), and the three-point statistic proposed in \citealt{zaldarriaga01b} (Appendix A) in a manner that is distinguishable from a power-law $T-\Delta_b$ relation.  Thus, the temperature fluctuations produced during a $z\sim3$ \HeII\ reionization by quasars would probably have evaded previous searches and also likely would evade future searches using these standard statistics.

Interestingly, we find that realistic models for large-scale temperature fluctuations could have a significant effect on the $3$D Ly$\alpha$ forest power spectrum.  This statistic can be measured by cross-correlating multiple quasar sightlines.  In the line-of-sight power spectrum, the aliasing of small-scale $3$D density power to larger scale line-of-sight modes dramatically suppresses the prominence of temperature fluctuations for physically motivated models.  However, we found that physically motivated temperature models could impart an order unity increase in the $3$D power at $k \sim 0.1~$Mpc$^{-1}$.  We showed that the impact of temperature fluctuations on the $3$D power spectrum could be understood with a simple analytic model. 

Intensity fluctuations could also alter the large-scale correlations in the Ly$\alpha$ forest.  These too would have been hidden in the line-of-sight power spectrum by aliasing effects.  At sufficiently large scales, intensity fluctuations will be the dominant source of fluctuations in the forest, and the fluctuations will become more prominent with increasing redshift.  The impact of intensity fluctuations on $P_F$ is likely to be distinguishable spectrally from the impact of temperature inhomogeneities.

%We quantified the sensitivity of surveys to the $3$D Ly$\alpha$ forest power spectrum.  We found that the sensitivity of a survey could be quantified with a single number, the effective density of the background sources.  
BOSS is forecasted to measure the $3$D flux power spectrum $P_F$ with percent-level accuracy at $z\approx 2.5$ in a $k$-space shell with $\Delta k \sim k$.  Such a measurement would constrain the effects discussed here, and we anticipate that these extra sources of fluctuations can be distinguished from other uncertainties in $P_F$ (such as in the cosmology) when their impact is larger than $\sim 10\%$.  In addition, BOSS is capable of placing interesting constraints on this statistic to redshifts as high as $z\approx 3$ and BigBOSS to $z\approx 4$.  Deeper surveys on an $8-10~$m class telescope can push this measurement to even higher redshifts \citep{mcquinnwhite}.  Furthermore, $P_F$ can be separated into different angular components with BOSS and future quasar surveys, which will facilitate the separation of the density, intensity, and temperature contributions.  The cross correlation of the Ly$\alpha$ forest with another tracer of large-scale structure (such as a high-z galaxy survey) could also enable this separation \citep{2011MNRAS.410.1130G, mcquinnwhite}.

The impact of temperature and intensity fluctuations will complicate attempts to constrain cosmic distances using the BAO features in the $3$D Ly$\alpha$ forest.  Because there is not one unique template for the spectrum of $T$ and $J$ fluctuations, marginalizing over their potential impact would likely require a fairly general parameterization for the continuum on which the BAO sits.  Such a marginalization procedure would reduce the sensitivity to cosmological parameters.  Temperature and intensity fluctuations may also impart (via their correlation with density) a scale dependence to the BAO amplitude. 

\citet{cen09} found that temperature fluctuations from models of \HI\ reionization had a larger effect on the line-of-sight Ly$\alpha$ forest power spectrum than we have found in our temperature models.  They examined $P_{F}^{\rm los}$ at $z=4$ and $z=5$, using simulations of hydrogen reionization.  They found that temperature fluctuations resulted in an increase in $P_{F}^{\rm los}$ by $5-10\%$ at $k \sim 10^{-2}~$s~km$^{-1}$ and $20-30\%$ at $k \sim 10^{-3}~$s~km$^{-1}$ -- the latter being roughly an order of magnitude larger than what our temperature models produce at $z=3$.  The larger effect found in \citet{cen09} likely owes to two reasons.  First, if the hydrogen were reionized by numerous dwarf galaxies as is assumed in the \citet{cen09} simulations, there would have been a stronger anti-correlation after this process completed between $T$ and the large-scale density at relevant scales than in our \HeII\ reionization simulations.  The effect of temperature fluctuations on $P_{F}^{\rm los}$ could be significantly enhanced by such an anti-correlation, as can be noted from equation~(\ref{eqn:Pklg2}).  Second, the density power decreases with increasing redshift, which results in the temperature fluctuations (which tend to have $\Delta T/T \sim 1$ because of the photoionization physics) becoming relatively larger.  The impact of temperature fluctuations in the \citet{cen09} models should be even more dramatic on the $3$D Ly$\alpha$ forest power spectrum.\\  
%.  It will be difficult to achieve the number densities of $10^{-3}~$Mpc$^{-3}$ on quasars required for measuring the $3$D power spectrum at high redshift, but such a technique can in principle be applied to galaxies as long as the signal to noise on the continuum of the entire Ly$\alpha$ forest region of the spectrum is greater than $10 \, n^{-1}$ for imaging.   {\bf [Fix this estimate]}  

%Thus, this study provides an additional motivation for what is the next frontier in Ly$\alpha$ forest analyses, the measurement of $3$D correlations in the Ly$\alpha$ forest.\\

We would like to thank Steven Furlanetto, Shirley Ho, Avi Loeb, and Nic Ross for useful discussions, and Martin White for useful comments on the manuscript.  We thank Claude-Andr{\'e} Faucher-Gigu{\`e}re for providing the hydrodynamic simulations used in some of our calculations.  MM is supported by the NASA Einstein Fellowship.

\bibliographystyle{mn2e}
\bibliography{References}

\begin{thebibliography}{}

\bibitem[\protect\citeauthoryear{{Abel} \& {Haehnelt}}{{Abel} \&
  {Haehnelt}}{1999}]{abel99}
{Abel} T.,  {Haehnelt} M.~G.,  1999, \apjl, 520, L13

\bibitem[\protect\citeauthoryear{{Agafonova}, {Levshakov}, {Reimers},
  {Fechner}, {Tytler}, {Simcoe} \& {Songaila}}{{Agafonova}
  et~al.}{2007}]{agafonova07}
{Agafonova} I.~I.,  {Levshakov} S.~A.,  {Reimers} D.,  {Fechner} C.,  {Tytler}
  D.,  {Simcoe} R.~A.,    {Songaila} A.,  2007, \aap, 461, 893

\bibitem[\protect\citeauthoryear{{Becker}, {Bolton}, {Haehnelt} \&
  {Sargent}}{{Becker} et~al.}{2011}]{becker10}
{Becker} G.~D.,  {Bolton} J.~S.,  {Haehnelt} M.~G.,    {Sargent} W.~L.~W.,
  2011, \mnras, 410, 1096

\bibitem[\protect\citeauthoryear{{Bolton}, {Becker}, {Wyithe}, {Haehnelt} \&
  {Sargent}}{{Bolton} et~al.}{2010}]{bolton10}
{Bolton} J.~S.,  {Becker} G.~D.,  {Wyithe} J.~S.~B.,  {Haehnelt} M.~G.,
  {Sargent} W.~L.~W.,  2010, \mnras, 406, 612

\bibitem[\protect\citeauthoryear{{Bolton}, {Viel}, {Kim}, {Haehnelt} \&
  {Carswell}}{{Bolton} et~al.}{2008}]{bolton08}
{Bolton} J.~S.,  {Viel} M.,  {Kim} T.,  {Haehnelt} M.~G.,    {Carswell} R.~F.,
  2008, \mnras, 386, 1131

\bibitem[\protect\citeauthoryear{{Cen}, {McDonald}, {Trac} \& {Loeb}}{{Cen}
  et~al.}{2009}]{cen09}
{Cen} R.,  {McDonald} P.,  {Trac} H.,    {Loeb} A.,  2009, \apjl, 706, L164

\bibitem[\protect\citeauthoryear{{Croft}}{{Croft}}{2004}]{croft04}
{Croft} R.~A.~C.,  2004, \apj, 610, 642

\bibitem[\protect\citeauthoryear{{Croft}, {Weinberg}, {Bolte}, {Burles},
  {Hernquist}, {Katz}, {Kirkman} \& {Tytler}}{{Croft} et~al.}{2002}]{croft02}
{Croft} R.~A.~C.,  {Weinberg} D.~H.,  {Bolte} M.,  {Burles} S.,  {Hernquist}
  L.,  {Katz} N.,  {Kirkman} D.,    {Tytler} D.,  2002, \apj, 581, 20

\bibitem[\protect\citeauthoryear{{Dav{\'e}}, {Hernquist}, {Katz} \&
  {Weinberg}}{{Dav{\'e}} et~al.}{1999}]{dave99}
{Dav{\'e}} R.,  {Hernquist} L.,  {Katz} N.,    {Weinberg} D.~H.,  1999, \apj,
  511, 521

\bibitem[\protect\citeauthoryear{{Davidsen}, {Kriss} \& {Zheng}}{{Davidsen}
  et~al.}{1996}]{davidsen96}
{Davidsen} A.~F.,  {Kriss} G.~A.,    {Zheng} W.,  1996, \nat, 380, 47

\bibitem[\protect\citeauthoryear{{Dixon} \& {Furlanetto}}{{Dixon} \&
  {Furlanetto}}{2009}]{dixonfurlanetto}
{Dixon} K.~L.,  {Furlanetto} S.~R.,  2009, \apj, 706, 970

\bibitem[\protect\citeauthoryear{{Fang} \& {White}}{{Fang} \&
  {White}}{2004}]{fang04}
{Fang} T.,  {White} M.,  2004, \apjl, 606, L9

\bibitem[\protect\citeauthoryear{{Faucher-Gigu{\`e}re}, {Lidz}, {Hernquist} \&
  {Zaldarriaga}}{{Faucher-Gigu{\`e}re} et~al.}{2008}]{faucher08}
{Faucher-Gigu{\`e}re} C.,  {Lidz} A.,  {Hernquist} L.,    {Zaldarriaga} M.,
  2008, \apjl, 682, L9

\bibitem[\protect\citeauthoryear{{Faucher-Gigu{\`e}re}, {Prochaska}, {Lidz},
  {Hernquist} \& {Zaldarriaga}}{{Faucher-Gigu{\`e}re} et~al.}{2008}]{faucher07}
{Faucher-Gigu{\`e}re} C.,  {Prochaska} J.~X.,  {Lidz} A.,  {Hernquist} L.,
  {Zaldarriaga} M.,  2008, \apj, 681, 831

\bibitem[\protect\citeauthoryear{{Furlanetto}}{{Furlanetto}}{2009}]{furlanettoJfluc}
{Furlanetto} S.~R.,  2009, \apj, 700, 1666

\bibitem[\protect\citeauthoryear{{Furlanetto} \& {Dixon}}{{Furlanetto} \&
  {Dixon}}{2010}]{furlanettodixon}
{Furlanetto} S.~R.,  {Dixon} K.~L.,  2010, \apj, 714, 355

\bibitem[\protect\citeauthoryear{{Furlanetto} \& {Oh}}{{Furlanetto} \&
  {Oh}}{2008a}]{furlanetto07b}
{Furlanetto} S.~R.,  {Oh} S.~P.,  2008a, \apj, 682, 14

\bibitem[\protect\citeauthoryear{{Furlanetto} \& {Oh}}{{Furlanetto} \&
  {Oh}}{2008b}]{furlanetto07a}
{Furlanetto} S.~R.,  {Oh} S.~P.,  2008b, \apj, 681, 1

\bibitem[\protect\citeauthoryear{{Gnedin} \& {Hui}}{{Gnedin} \&
  {Hui}}{1998}]{gnedin98}
{Gnedin} N.~Y.,  {Hui} L.,  1998, \mnras, 296, 44

\bibitem[\protect\citeauthoryear{{Guha Sarkar}, {Bharadwaj}, {Choudhury} \&
  {Datta}}{{Guha Sarkar} et~al.}{2011}]{2011MNRAS.410.1130G}
{Guha Sarkar} T.,  {Bharadwaj} S.,  {Choudhury} T.~R.,    {Datta} K.~K.,  2011,
  \mnras, 410, 1130

\bibitem[\protect\citeauthoryear{{Hernquist}, {Katz}, {Weinberg} \&
  {Miralda-Escud{\'e}}}{{Hernquist} et~al.}{1996}]{hernquist96}
{Hernquist} L.,  {Katz} N.,  {Weinberg} D.~H.,    {Miralda-Escud{\'e}} J.,
  1996, \apjl, 457, L51+

\bibitem[\protect\citeauthoryear{{Hopkins}, {Richards} \&
  {Hernquist}}{{Hopkins} et~al.}{2007}]{hopkins07a}
{Hopkins} P.~F.,  {Richards} G.~T.,    {Hernquist} L.,  2007, \apj, 654, 731

\bibitem[\protect\citeauthoryear{{Hui} \& {Gnedin}}{{Hui} \&
  {Gnedin}}{1997}]{hui97}
{Hui} L.,  {Gnedin} N.~Y.,  1997, \mnras, 292, 27

\bibitem[\protect\citeauthoryear{{Hui} \& {Haiman}}{{Hui} \&
  {Haiman}}{2003}]{hui03}
{Hui} L.,  {Haiman} Z.,  2003, \apj, 596, 9

\bibitem[\protect\citeauthoryear{{Kaiser}}{{Kaiser}}{1987}]{kaiser87}
{Kaiser} N.,  1987, \mnras, 227, 1

\bibitem[\protect\citeauthoryear{{Katz}, {Weinberg}, {Hernquist} \&
  {Miralda-Escude}}{{Katz} et~al.}{1996}]{katz96}
{Katz} N.,  {Weinberg} D.~H.,  {Hernquist} L.,    {Miralda-Escude} J.,  1996,
  \apjl, 457, L57+

\bibitem[\protect\citeauthoryear{{Kim}, {Bolton}, {Viel}, {Haehnelt} \&
  {Carswell}}{{Kim} et~al.}{2007}]{kim07}
{Kim} T.,  {Bolton} J.~S.,  {Viel} M.,  {Haehnelt} M.~G.,    {Carswell} R.~F.,
  2007, \mnras, 382, 1657

\bibitem[\protect\citeauthoryear{{Komatsu} et~al.,}{{Komatsu}
  et~al.}{2011}]{komatsu10}
{Komatsu} E.,  et~al., 2011, \apjs, 192, 18

\bibitem[\protect\citeauthoryear{{Lai}, {Lidz}, {Hernquist} \&
  {Zaldarriaga}}{{Lai} et~al.}{2006}]{lai06}
{Lai} K.,  {Lidz} A.,  {Hernquist} L.,    {Zaldarriaga} M.,  2006, \apj, 644,
  61

\bibitem[\protect\citeauthoryear{{Lee}}{{Lee}}{2011}]{2011arXiv1103.2780L}
{Lee} K.,  2011, ArXiv e-prints

\bibitem[\protect\citeauthoryear{{Lee} \& {Spergel}}{{Lee} \&
  {Spergel}}{2010}]{lee10}
{Lee} K.,  {Spergel} D.~N.,  2010, arxiv:1007.3734

\bibitem[\protect\citeauthoryear{{Lidz}, {Faucher-Gigu{\`e}re}, {Dall'Aglio},
  {McQuinn}, {Fechner}, {Zaldarriaga}, {Hernquist} \& {Dutta}}{{Lidz}
  et~al.}{2010}]{lidz09}
{Lidz} A.,  {Faucher-Gigu{\`e}re} C.,  {Dall'Aglio} A.,  {McQuinn} M.,
  {Fechner} C.,  {Zaldarriaga} M.,  {Hernquist} L.,    {Dutta} S.,  2010, \apj,
  718, 199

\bibitem[\protect\citeauthoryear{{McDonald}}{{McDonald}}{2003}]{mcdonald03}
{McDonald} P.,  2003, \apj, 585, 34

\bibitem[\protect\citeauthoryear{{McDonald} et~al.,}{{McDonald}
  et~al.}{2005}]{mcdonald05b}
{McDonald} P.,  et~al., 2005, \apj, 635, 761

\bibitem[\protect\citeauthoryear{{McDonald}, {Miralda-Escud{\'e}}, {Rauch},
  {Sargent}, {Barlow} \& {Cen}}{{McDonald} et~al.}{2001}]{mcdonald01b}
{McDonald} P.,  {Miralda-Escud{\'e}} J.,  {Rauch} M.,  {Sargent} W.~L.~W.,
  {Barlow} T.~A.,    {Cen} R.,  2001, \apj, 562, 52

\bibitem[\protect\citeauthoryear{{McDonald}, {Miralda-Escud{\'e}}, {Rauch},
  {Sargent}, {Barlow}, {Cen} \& {Ostriker}}{{McDonald}
  et~al.}{2000}]{mcdonald00}
{McDonald} P.,  {Miralda-Escud{\'e}} J.,  {Rauch} M.,  {Sargent} W.~L.~W.,
  {Barlow} T.~A.,  {Cen} R.,    {Ostriker} J.~P.,  2000, \apj, 543, 1

\bibitem[\protect\citeauthoryear{{McDonald}, {Seljak}, {Cen}, {Bode} \&
  {Ostriker}}{{McDonald} et~al.}{2005}]{mcdonald05}
{McDonald} P.,  {Seljak} U.,  {Cen} R.,  {Bode} P.,    {Ostriker} J.~P.,  2005,
  \mnras, 360, 1471

\bibitem[\protect\citeauthoryear{{McQuinn}}{{McQuinn}}{2009}]{mcquinnGP}
{McQuinn} M.,  2009, \apjl, 704, L89

\bibitem[\protect\citeauthoryear{{McQuinn}, {Lidz}, {Zaldarriaga}, {Hernquist},
  {Hopkins}, {Dutta} \& {Faucher-Gigu{\`e}re}}{{McQuinn}
  et~al.}{2009}]{mcquinn09}
{McQuinn} M.,  {Lidz} A.,  {Zaldarriaga} M.,  {Hernquist} L.,  {Hopkins} P.~F.,
   {Dutta} S.,    {Faucher-Gigu{\`e}re} C.,  2009, \apj, 694, 842

\bibitem[\protect\citeauthoryear{{McQuinn} \& {White}}{{McQuinn} \&
  {White}}{2011}]{mcquinnwhite}
{McQuinn} M.,  {White} M.,  2011, ArXiv e-prints

\bibitem[\protect\citeauthoryear{{Meiksin}}{{Meiksin}}{2000}]{meiksin00}
{Meiksin} A.,  2000, \mnras, 314, 566

\bibitem[\protect\citeauthoryear{{Meiksin} \& {White}}{{Meiksin} \&
  {White}}{2004}]{meiksin04}
{Meiksin} A.,  {White} M.,  2004, \mnras, 350, 1107

\bibitem[\protect\citeauthoryear{{Mesinger} \& {Furlanetto}}{{Mesinger} \&
  {Furlanetto}}{2009}]{2009MNRAS.400.1461M}
{Mesinger} A.,  {Furlanetto} S.,  2009, \mnras, 400, 1461

\bibitem[\protect\citeauthoryear{{Miralda-Escud{\' e}}, {Haehnelt} \&
  {Rees}}{{Miralda-Escud{\' e}} et~al.}{2000}]{miralda00}
{Miralda-Escud{\' e}} J.,  {Haehnelt} M.,    {Rees} M.~J.,  2000, \apj, 530, 1

\bibitem[\protect\citeauthoryear{{Miralda-Escud{\'e}}, {Cen}, {Ostriker} \&
  {Rauch}}{{Miralda-Escud{\'e}} et~al.}{1996}]{miralda96}
{Miralda-Escud{\'e}} J.,  {Cen} R.,  {Ostriker} J.~P.,    {Rauch} M.,  1996,
  \apj, 471, 582

\bibitem[\protect\citeauthoryear{{Miralda-Escud{\'e}} \&
  {Rees}}{{Miralda-Escud{\'e}} \& {Rees}}{1994}]{miralda94}
{Miralda-Escud{\'e}} J.,  {Rees} M.~J.,  1994, \mnras, 266, 343

\bibitem[\protect\citeauthoryear{{Peacock} \& {Dodds}}{{Peacock} \&
  {Dodds}}{1996}]{peacock96}
{Peacock} J.~A.,  {Dodds} S.~J.,  1996, \mnras, 280, L19

\bibitem[\protect\citeauthoryear{{Peeples}, {Weinberg}, {Dav{\'e}}, {Fardal} \&
  {Katz}}{{Peeples} et~al.}{2010}]{peeples09a}
{Peeples} M.~S.,  {Weinberg} D.~H.,  {Dav{\'e}} R.,  {Fardal} M.~A.,    {Katz}
  N.,  2010, \mnras, 404, 1281

\bibitem[\protect\citeauthoryear{{Prochaska}, {Worseck} \&
  {O'Meara}}{{Prochaska} et~al.}{2009}]{prochaska09}
{Prochaska} J.~X.,  {Worseck} G.,    {O'Meara} J.~M.,  2009, \apjl, 705, L113

\bibitem[\protect\citeauthoryear{{Ricotti}, {Gnedin} \& {Shull}}{{Ricotti}
  et~al.}{2000}]{2000ApJ...534...41R}
{Ricotti} M.,  {Gnedin} N.~Y.,    {Shull} J.~M.,  2000, \apj, 534, 41

\bibitem[\protect\citeauthoryear{{Schaye}, {Theuns}, {Rauch}, {Efstathiou} \&
  {Sargent}}{{Schaye} et~al.}{2000}]{2000MNRAS.318..817S}
{Schaye} J.,  {Theuns} T.,  {Rauch} M.,  {Efstathiou} G.,    {Sargent}
  W.~L.~W.,  2000, \mnras, 318, 817

\bibitem[\protect\citeauthoryear{{Shull}, {France}, {Danforth}, {Smith} \&
  {Tumlinson}}{{Shull} et~al.}{2010}]{shull10}
{Shull} M.,  {France} K.,  {Danforth} C.,  {Smith} B.,    {Tumlinson} J.,
  2010, ArXiv e-prints

\bibitem[\protect\citeauthoryear{{Slosar}, {Ho}, {White} \& {Louis}}{{Slosar}
  et~al.}{2009}]{slosar09}
{Slosar} A.,  {Ho} S.,  {White} M.,    {Louis} T.,  2009, Journal of Cosmology
  and Astro-Particle Physics, 10, 19

\bibitem[\protect\citeauthoryear{{Springel}}{{Springel}}{2005}]{springel05}
{Springel} V.,  2005, \mnras, 364, 1105

\bibitem[\protect\citeauthoryear{{Theuns}, {Schaye} \& {Haehnelt}}{{Theuns}
  et~al.}{2000}]{2000MNRAS.315..600T}
{Theuns} T.,  {Schaye} J.,    {Haehnelt} M.~G.,  2000, \mnras, 315, 600

\bibitem[\protect\citeauthoryear{{Theuns}, {Schaye}, {Zaroubi}, {Kim},
  {Tzanavaris} \& {Carswell}}{{Theuns} et~al.}{2002}]{theunsschaye02}
{Theuns} T.,  {Schaye} J.,  {Zaroubi} S.,  {Kim} T.,  {Tzanavaris} P.,
  {Carswell} B.,  2002, \apjl, 567, L103

\bibitem[\protect\citeauthoryear{{Theuns} \& {Zaroubi}}{{Theuns} \&
  {Zaroubi}}{2000}]{theuns00}
{Theuns} T.,  {Zaroubi} S.,  2000, \mnras, 317, 989

\bibitem[\protect\citeauthoryear{{Theuns}, {Zaroubi}, {Kim}, {Tzanavaris} \&
  {Carswell}}{{Theuns} et~al.}{2002}]{2002MNRAS.332..367T}
{Theuns} T.,  {Zaroubi} S.,  {Kim} T.-S.,  {Tzanavaris} P.,    {Carswell}
  R.~F.,  2002, \mnras, 332, 367

\bibitem[\protect\citeauthoryear{{Trac}, {Cen} \& {Loeb}}{{Trac}
  et~al.}{2008}]{trac08}
{Trac} H.,  {Cen} R.,    {Loeb} A.,  2008, \apjl, 689, L81

\bibitem[\protect\citeauthoryear{{Viel}, {Bolton} \& {Haehnelt}}{{Viel}
  et~al.}{2009}]{2009MNRAS.tmpL.290V}
{Viel} M.,  {Bolton} J.~S.,    {Haehnelt} M.~G.,  2009, \mnras, pp L290+

\bibitem[\protect\citeauthoryear{{White}, {Pope}, {Carlson}, {Heitmann},
  {Habib}, {Fasel}, {Daniel} \& {Lukic}}{{White} et~al.}{2010}]{white10}
{White} M.,  {Pope} A.,  {Carlson} J.,  {Heitmann} K.,  {Habib} S.,  {Fasel}
  P.,  {Daniel} D.,    {Lukic} Z.,  2010, \apj, 713, 383

\bibitem[\protect\citeauthoryear{{Zaldarriaga}}{{Zaldarriaga}}{2002}]{zaldarriaga02}
{Zaldarriaga} M.,  2002, \apj, 564, 153

\bibitem[\protect\citeauthoryear{{Zaldarriaga}, {Hui} \&
  {Tegmark}}{{Zaldarriaga} et~al.}{2001}]{zaldarriaga01c}
{Zaldarriaga} M.,  {Hui} L.,    {Tegmark} M.,  2001, \apj, 557, 519

\bibitem[\protect\citeauthoryear{{Zaldarriaga}, {Seljak} \&
  {Hui}}{{Zaldarriaga} et~al.}{2001}]{zaldarriaga01b}
{Zaldarriaga} M.,  {Seljak} U.,    {Hui} L.,  2001, \apj, 551, 48

\bibitem[\protect\citeauthoryear{{Zuo}}{{Zuo}}{1992a}]{zuo92}
{Zuo} L.,  1992a, \mnras, 258, 36

\bibitem[\protect\citeauthoryear{{Zuo}}{{Zuo}}{1992b}]{zuo92b}
{Zuo} L.,  1992b, \mnras, 258, 45

\end{thebibliography}

\appendix

\section{Other Line-of-Sight Statistics}

\subsection{Wavelets}
\label{ss:wavelet}

Several studies have suggested using wavelet functions to search for temperature fluctuations in the Ly$\alpha$ forest \citep{meiksin00, 2002MNRAS.332..367T, zaldarriaga02}.  This method uses the wavelet property that they are localized in configuration in addition to Fourier space.
  The idea is to convolve the Ly$\alpha$ forest transmission field with a wavelet that is sensitive to the amount of small-scale power to search for spatial variations in the power and, thus, temperature fluctuations.  

\begin{figure}
\rotatebox{-90}{\epsfig{file=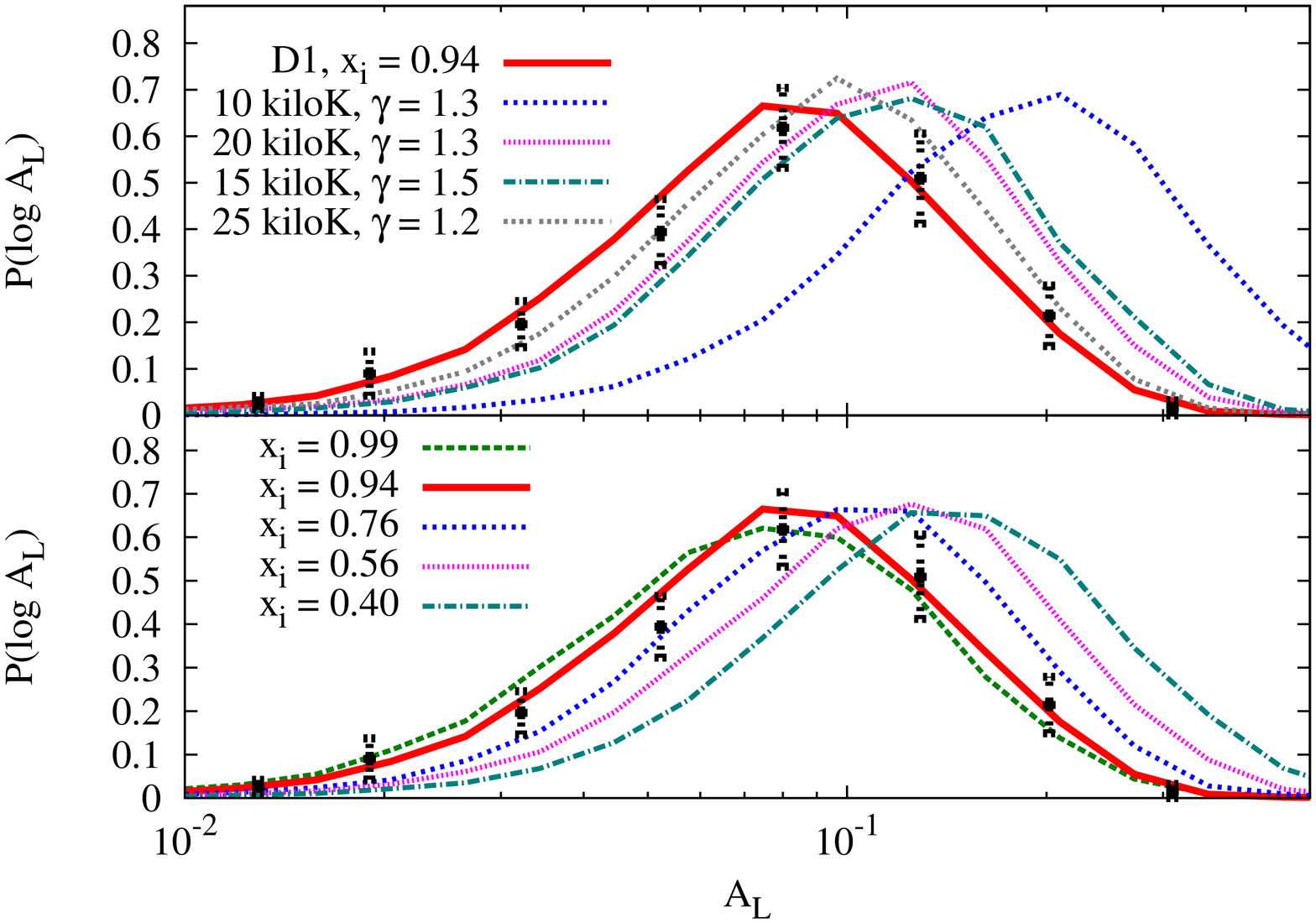, height=8.5cm}}
\caption{PDF of the wavelet coefficients $A_L$ with $L=500~$km~s$^{-1}$ for several temperature models.  The points with errorbars are from the $z=3$ measurement of \citet{lidz09} and the solid red curves represent the $z=3$ snapshot in the simulation.
\label{fig:wavelet}}
\end{figure}

\begin{figure}
\rotatebox{-90}{\epsfig{file=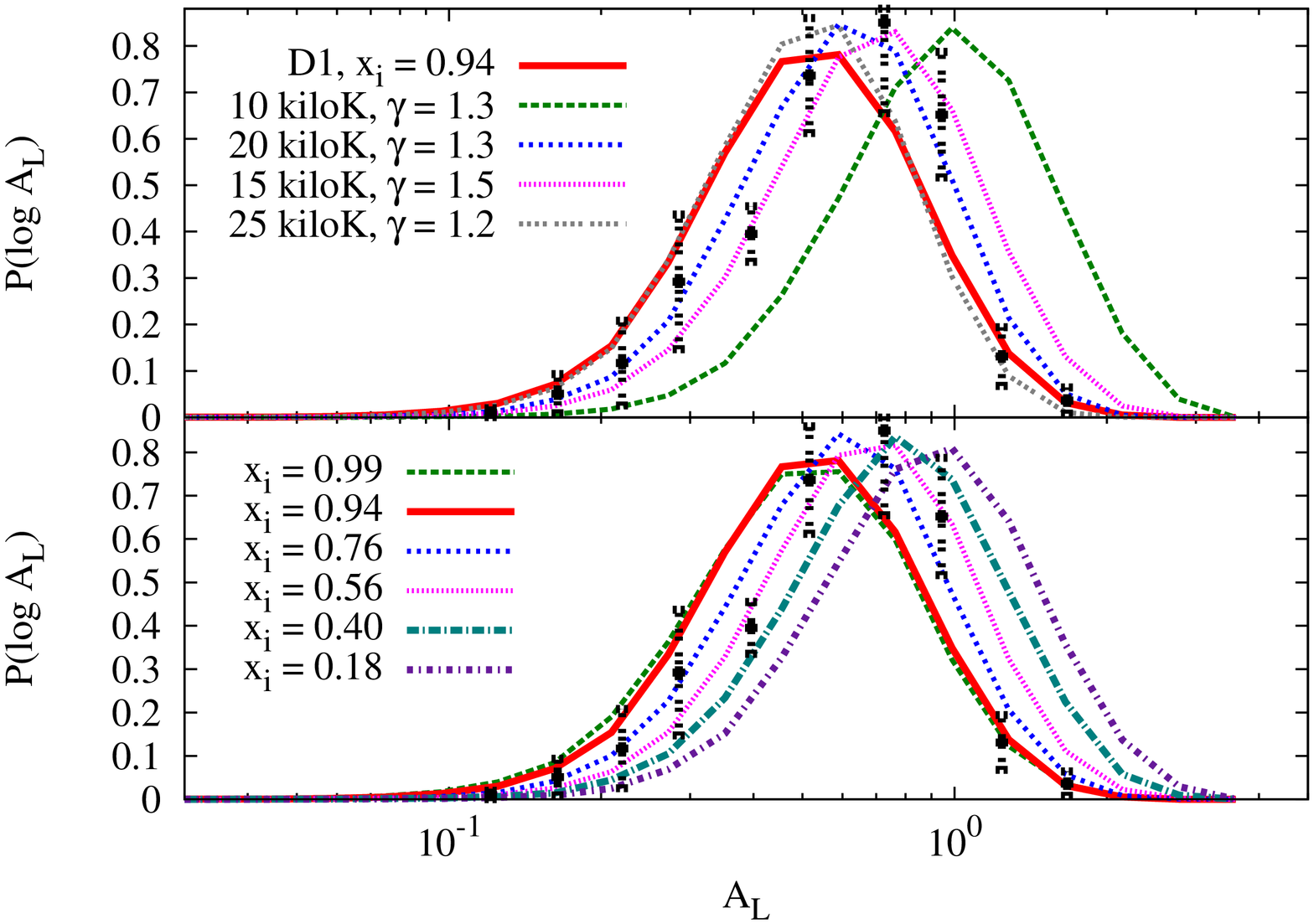, height=8.5cm}}
\caption{Same as Figure \ref{fig:wavelet} but for $z=4.2$.  The teal curve in the bottom panel represents the $z=4$ snapshot in the simulation. \label{fig:waveletZ4}}
\end{figure}

We compare the wavelet predictions of different temperature models with the recent measurement by \citet{lidz09}.  This study applied this technique to $40$ high resolution, high S/N Ly$\alpha$ forest spectra spanning $2 \lesssim z  \lesssim 4.5$.  The wavelet filter that \citet{lidz09} used is
\begin{equation}
\Psi(x) = C\, \exp \left({\it i} k_0 x \right) \; \exp \left[-\frac{x^2}{2 \,s_n^2} \right]. 
\end{equation}
\citet{lidz09} chose $C$ such that $\Sigma_{i=1}^N \Delta x \, \Psi(x)^2/N = 1$ for $\Delta x = 4.4~$km~s$^{-1}$.  \citet{lidz09} found $s_n = 69.7$~km~s$^{-1}$ and $k_0 \, s_n = 6$ to be a good compromise between maximizing sensitivity to temperature while minimizing the impact of instrumental noise.   Conveniently, the Fourier transform of $\Psi(x)$, which we denote as $\tilde{\Psi}(k)$, is a Gaussian centered around $ k = k_0$ with standard deviation $s_n^{-1}$.   The full width half maximum of $\tilde{\Psi}(k)$ spans $0.06 < k < 0.1~$s~km$^{-1}$, and $\tilde{\Psi}(k)$ is plotted in the top panels in Figures \ref{fig:pk} and \ref{fig:pk2} (thin curves, arbitrarily normalized).  The wavelet filter that was used in other wavelet studies of the forest is qualitatively similar to the \citet{lidz09} filter, and we expect our conclusions are robust to the exact filter choice (such as the curvature statistic of \citealt{becker10}).
% resulting in $\pm 1$ standard deviation on $\tilde{\Psi}(k)$ spanning $0.07 \lesssim k \lesssim 0.1~$km~s$^{-1}$ (on the exponential tail of $P_{\delta_F}$).

\citet{lidz09} primarily analyzed the PDF of 
\begin{equation}
A_L (x) = \frac{1}{2\, L} \int_{-L}^{L} dx \, \left | \Psi(x) \circ \delta_F(x) \right |^2,
\end{equation}
where $``\circ"$ denotes a convolution, and the integral averages the convolved signal over $L = 500~$km~s$^{-1}$ in order to reduce the noise.  Thus, their $A_L$ was the average of the wavelet power in a $\approx 10~$Mpc region.  The mean of this PDF is sensitive to the average temperature (because it is a measure of the average power within the wavelet bandpass), whereas the width is a measure of the spatial variance in the temperature.  However, cosmic variance in the forest is the primary determinant of the width of this PDF, and temperature fluctuations would manifest as an excess in the width over what is expected from simple models for the IGM thermal state.

%The mean of this PDF is equal to $\int dk/{2\pi} \, |\tilde{\Psi}(k)|^2 P_{\cal F}$ as $L\rightarrow 0$.  {\bf must be better way to right this}  

Figures \ref{fig:wavelet} and \ref{fig:waveletZ4} plot the predictions for the wavelet PDF at $z=3$ and $z=4.2$ using different temperature models.  These curves were calculated from the same simulation skewers as the $P_F^{\rm los}$ curves in Figures \ref{fig:pk} and \ref{fig:pk2}.  The points with error bars are the measured values from \citet{lidz09}.  The top panel in both figures explore toy power-law $T-\Delta_b$ models.  (See \citet{lidz09} for a more extensive comparison of such models.)  

The wavelet PDF of the $\gamma = 1.3$, $T_0 = 10~$kiloK model is quite discrepant with that of the $\gamma = 1.3$, $20~$kiloK model (Fig. \ref{fig:wavelet}).  The data favors the latter of these two models, and would also be inconsistent with a $50\%$ mix of both temperatures (which would be the average of these two curves).  Section \ref{sec:background} suggested that a more realistic toy fluctuating-temperature model is half the volume with $T_0 = 15~$kiloK and $\gamma = 1.5$ and the other half with $T_0 = 25~$kiloK  and $\gamma = 1.2$.  Unfortunately, the top panel illustrates that $T_0 = 15~$kiloK and $\gamma = 1.5$ produces a very similar wavelet PDF to the $T_0 = 25~$kiloK and $\gamma = 1.2$ case.  Thus, as we found for $P_F^{\rm los}$ in Section \ref{ss:sslos}, a $50\%$ mix of these two models would be difficult to distinguish from a single $T_0$ and $\gamma$ model.

The red solid curves in Figures \ref{fig:wavelet} and \ref{fig:waveletZ4} represent the simulation result near the end of \HeII\ reionization.  These curves have a very similar width to the other curves in their respective panel despite the fact that these curves include  dispersion in the temperature.  %The lack of broadening can be understood by the fact that the colder regions in the simulations have steeper $\gamma$ than the hotter one, and the toy case curves in the top panel illustrate that this effect suppresses the broadening of the PDF.   
The bottom panels show the wavelet PDF at different times during the simulation.  Interestingly, none of these PDFs are noticeably broader than the power-law $T-\Delta_b$ case.  Over the course of \HeII\ reionization in the simulation, the mean of the PDF shifts to smaller values owing to the heating of the simulated IGM.  The teal and red curves represent the $z=3$ and $z=4$ outputs in the simulation.  The mean of the PDF in the simulation appears to evolve slightly more between these redshifts than the data, consistent with what we found in Section \ref{ss:sslos}.

%One might expect that temperature fluctuations become easier to detect in the wavelet PDF as the redshift increases:  The PDF becomes sensitive to the temperature in lower density regions with increasing redshift, regions where the simulations suggest that the temperature fluctuations caused by \HeII\ reionization should be larger.  However, we find temperature fluctuations are not significantly easier to detect in the PDF at $z=4.2$ compared to $z=3$.  This results partly because thermal broadening is less important in these lower density regions, which reduces the contrast between the wavelet PDF of the different temperature models.  Another difficulty with the wavelet PDF at $z\gtrsim4$ is that the interpretation is complicated by uncertainties in the mean flux, as discussed in \citet{lidz09}.   

In conclusion, measurements of the wavelet PDF constrain the mean temperature and place limits on temperature fluctuations at the level of $\Delta T/T \approx 1$ for slightly overdense gas.  However, temperature fluctuations at the smaller level found in the simulations of \citealt{mcquinn09} do not significantly effect the width of the wavelet PDF and would be extremely difficult to detect with this statistic.

%These results are not so unexpanded.  Since the small-scale power between hot and cold regions is not significantly different in our simulations (because of the different $\gamma$), this argues that the variance of the wavelet is not substantially affected compared to simple models for the heating of the IGM.  Figure \ref{fig:wavelet} confirms this suspicion, where the wavelet PDF for the same cases as in Figure \ref{fig:pk} are shown.  

%The mean of the wavelet PDF is primarily a measure of the power at small scales and, thus, the mean temperature.  The variance of the PDF is a measure of the power variations between regions of size $L$ and, thus, of temperature fluctuations.  

%\begin{figure}
%\rotatebox{-90}{\epsfig{file=power_spectrum_z3.eps, height=8.5cm}}
%\rotatebox{-90}{\epsfig{file=power_spectrum_z4.eps, height=8.5cm}}
%\caption{The power spectrum of the flux from the simulations and from models with a perfect $T-\Delta_b$ relationship.
%\label{fig:pkold}}
%\end{figure}

\subsection{Small-Scale Power -- Large-Scale Flux Correlation}
\label{sec:3point}

\begin{figure}
\rotatebox{-90}{\epsfig{file=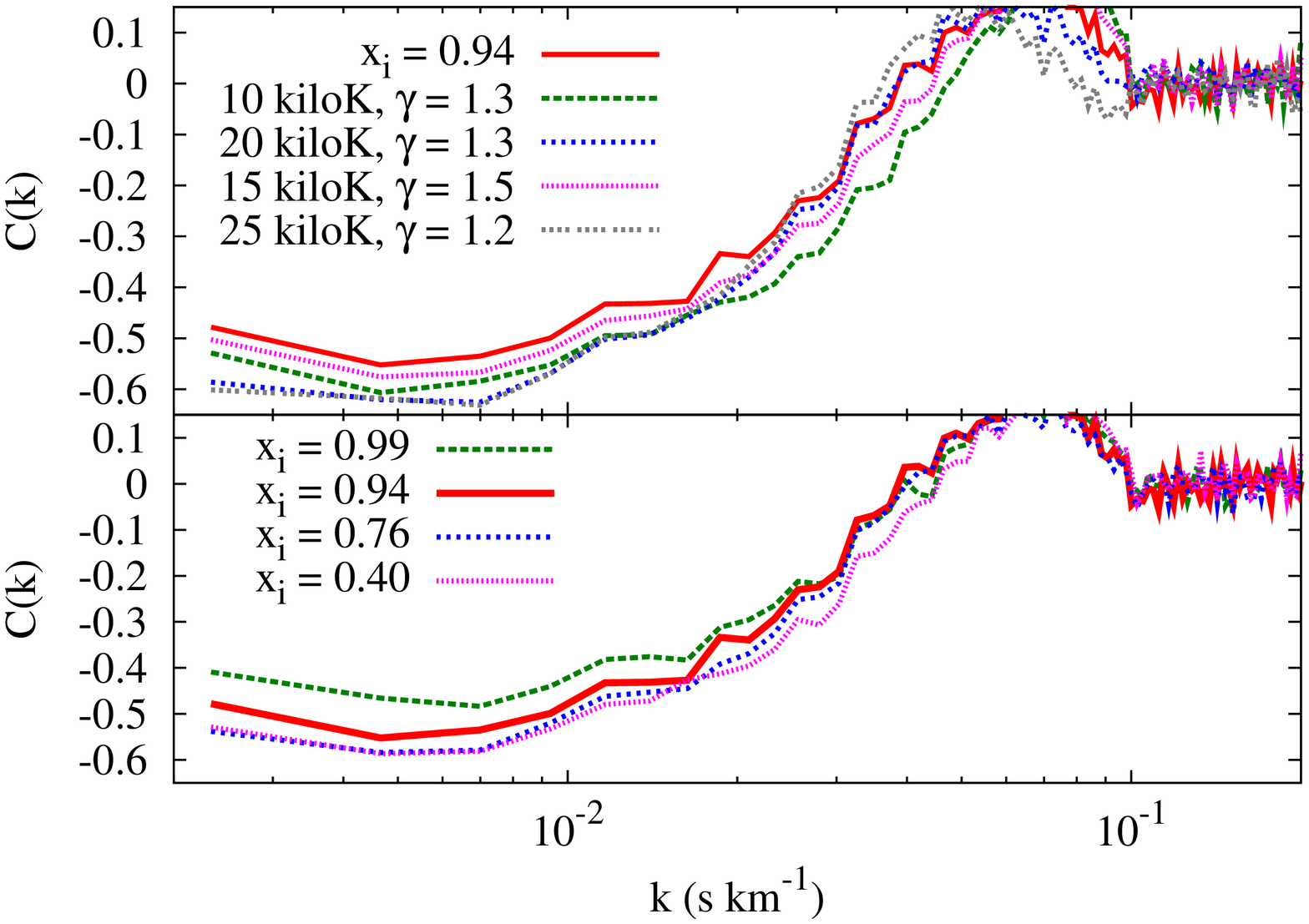, height=9cm}}
\caption{Three-point statistic outlined in Section \ref{sec:3point}, with $k_1 = 0.1$ and $k_2 = 0.2~$s~km$^{-1}$.  The top panel shows this statistic for different power-law $T-\Delta_b$ models and for the temperature in the simulation snapshot with $x_{\rm HeIII} = 0.94$. The bottom panel shows this statistic using the temperature from several snapshots of the \HeII\ reionization simulation.
\label{fig:zald3piont}}
\end{figure}

\citet{zaldarriaga01b} and \citet{fang04} discussed a particular three-point statistic that they argued provides an excellent test of the paradigm that gravitational instability shapes the transmission fluctuations in the Ly$\alpha$ forest.  They advocated the statistic 
\begin{equation}
C(k) \equiv \frac{P_{h \,\delta_F} (k)} {[P_{h}(k) \, P_{\delta_F} (k)]^{1/2}},
\end{equation}
where $P_{\delta_F}(k)$ is the power-spectrum of ${\delta}_F$, $P_{h}(k)$ is this of ${h}$, and $P_{h \delta_F}(k)$ is the cross power between ${h}$ and ${\delta}_F$.  Here, $h(x) \equiv \delta_H(x)^2$ and $\tilde{\delta}_H \equiv \tilde{\delta}_F(k) W_{k_1, k_2}(k)$, where $W_{k_1, k_2}(k)$ is a band-pass filter that transmits at $k_1 < k < k_2$ and tildes distinguish the Fourier dual.  Thus, $h$ is the square of the band pass-filtered flux field, making it a measure of the bandpass power.  This statistic is normalized such that a perfect correlation between $h$ with $\delta_F$ yields $C =1$.  \citet{fang04} showed that this statistic could place strong constraints on the level of temperature fluctuations.

As in \citet{zaldarriaga01b} and \citet{fang04}, we take $W_{k_1, k_2}(k)$ to be unity for $k_1 < k < k_2$ and zero otherwise.  The statistic $C(k)$ is a measure of the correlation between ${\delta}_F$ and the small-scale power in the high pass-filtered Ly$\alpha$ transmission field.  If gravity dominates the fluctuations in the forest, there will be more structure in a large-scale overdense (opaque) region, driving $C(k)$ negative.  Predictions for $C(k)$ are shown in the top panel of Figure \ref{fig:zald3piont} for different power-law $T-\Delta_b$ models.  The curves are calculated from the $1000$, $25/h~$Mpc random skewers drawn from the $z=3$ snapshot of the hydrodynamic simulation and with $k_1 = 0.1$ and $k_2 = 0.2~$s~km$^{-1}$. These curves all have $-0.6 < C < -0.5$ at $k \lesssim 0.01~$km~s$^{-1}$, and the curves with the largest $\gamma$ have slightly smaller $|C|$.  This trend results because, the larger the deviation from an isothermal relation, the more decorrelated a large-scale density mode is from the small-scale density power \citep{zaldarriaga01b}.

% We find that the difference in $C$ between different temperature models becomes smaller as the values of $k_1$ and $k_2$ are decreased.
%\footnote{The statistic $C(k)$ may be more robust to these systematics than the wavelet statistics used in \citet{lidz09}.  The only term in $C(k)$ that is biased by a systematic that increases the small-scale power is $P_{h}(k)$.  We find that dropping this term from $C$ has only a minor effect on the trends seen in Figure \ref{fig:zald3piont} for curves with similar mean temperatures. %, which can be understood from noting that this term is primarily a measure of the small-scale power and we found that the considered models had little effect on the small-scale power spectrum.  Therefore, we suggest that analysis attempts should drop this factor in their measurement.
%}  

Temperature fluctuations from reionization could also decorrelate the small-scale density power from the large-scale flux.  However, temperature fluctuations could also enhance the negative correlation because a large-scale hot region has increased transmission and also less small-scale power.  We find that the former effect is most important in our reionization simulations:  These temperature fluctuations decrease $|C|$, but not significantly. The red solid curve in the top panel of Figure \ref{fig:zald3piont} uses the temperature fluctuations from the $x_{\rm HeIII} = 0.94$ snapshot, and we find that $|C|$ is slightly smaller for this case compared to the power-law $T-\Delta_b$ models.  The bottom panel in Figure \ref{fig:zald3piont} plots $C$ using the temperature field from different snapshots of the \HeII\ reionization simulation.

\citet{fang04} showed that adding a lognormal dispersion at the Jeans scale around the mean $T-\Delta_b$ relation with standard deviation $0.2$ results in $|C|$ becoming significantly smaller.  They used this result to constrain the standard deviation in this relation to be $< 0.2$ from a single quasar sightline.  The temperature fluctuations in our \HeII\ reionization simulations have a standard deviation of $\sigma_T \approx 0.1$ (Fig. \ref{fig:Tdeltadist}), but the simulations' temperature fluctuations are correlated over $10$s of Mpc: In a large-scale hot or cold region, the dispersion in $T-\Delta_b$ is much smaller (Fig. \ref{fig:Tdelta_local}).  Likely because of these large-scale correlations, we find a significantly smaller suppression than in \citet{fang04}.  \citet{fang04} found the most dramatic effect for a higher pass filter than used in this section, in particular with $k_1 = 0.2$ and $k_2 = 0.3~$s~km$^{-1}$.  We find that if we use this filter choice, $|C|$ is reduced by a larger factor than for our fiducial filter (to a value as low as $-0.3$) but never to zero.  However, even the choice $k_1 = 0.1$ and $k_2 = 0.2~$s~km$^{-1}$ is pushing the limit on what can be applied to even the highest quality Ly$\alpha$ forest data \citep{lidz09}.%\footnote{Our predictions from the simulations may underestimate the amount of suppression.  In order to include the temperature fluctuations in the high resolution hydrodynamic simulation, we fit locally to the $T-\Delta_b$ relation in the simulations in the manner discussed in Section \ref{sec:methodology}.  There was always dispersion with standard deviation $\approx 0.1$ around our best-fit model that was lost in this process.  While is likely that some of this dispersion owes to noise in the number of rays that reached each cell, some of this dispersion is physical (reflecting that the attenuation of rays is modulated by Jeans scale clumps).  Therefore, our predictions for the suppression of $|C(k)|$ from the simulations may in fact be lower bounds.}

%Both \citet{zaldarriaga01b} and \citet{fang04} applied this statistic to a single quasar spectrum.  Because of their limited data set and because they calculated this statistic in one redshift bin that spanned the entire forest in their quasar, we are not able to compare our predictions with their measurement.   Even though this statistic is not extremely sensitive to our temperature fluctuation models, it may be worthwhile to measure it from a larger sample of spectra.  

\section{Peculiar Velocities}
\label{app:lognormal}

\begin{figure}
\rotatebox{-90}{\epsfig{file=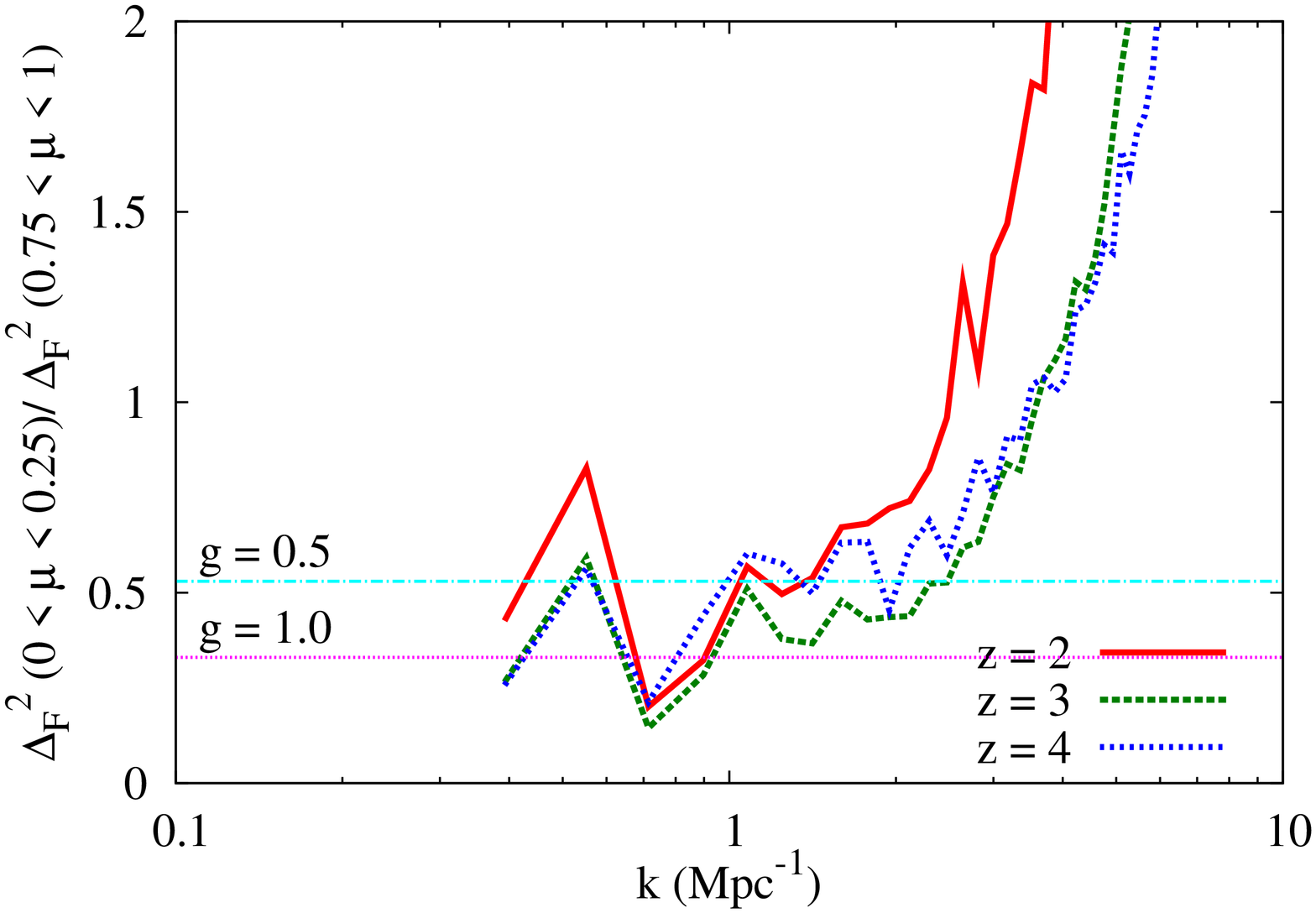, height=9cm}}
\caption{Ratio of the flux power spectrum for modes with $0 < \mu < 0.25$ to those with $0.75 < \mu < 1.0$.  The curves are measured from the $25/h~$Mpc simulation at $z=2, 3$ and $4$, assuming $\gamma = 1.3$.  The increase in these curves at small-scales owes to thermal broadening.  The approximate plateau in the ratio provides a measure for $g$.  The horizontal lines are the expected large-scale anisotropy for $g=0.5$ and $g=1.0$.  A larger box size would be required for a more precise determination.
\label{fig:pkani}}
\end{figure}

This Appendix quantifies the impact of peculiar velocities on large-scale flux correlations.  Peculiar velocities have two effects:  (1) The large-scale peculiar velocity field results in a redshift-space compression so that more systems appear in regions with converging flows. (2) The nonlinear peculiar velocity compresses or dilates Jeans-scale dense regions in absorption space (and dilates Jeans-scale voids).   On large scales, the latter effect enters by altering the bias of the forest.  The former effect produces the redshift-space anisotropy of this signal.  It affects $\tau_{\rm Ly\alpha}$ via the factor $(H \, a + dv/dx)^{-1}$, where $v$ and $x$ are the peculiar velocity and comoving distance along the line of sight.  In Fourier space this factor is simpler and approximately equal to $1+ \Omega_m(z)^{0.6} \mu^2 \delta$, where $\mu = \hat{k} \cdot \hat{n}$ \citep{kaiser87}. Thus, on linear scales the redshift-space power spectrum of $\delta_F$ is likely to have the form 
\begin{equation}
P_{F} \approx b^2 \left[ G^2 \, P_\Delta + 2\, G \, \epsilon^{-1} \, P_{\Delta X}  +  \epsilon^{-2} \, P_{X}  \right],
\label{eqn:Pklgnormal_largescales}
\end{equation}
where $G = (1 + g \mu^2)$, $g$ is the large-scale bias of velocity fluctuations, and $X$ is a placeholder for $T$ or $-J$ fluctuations.  The average of $G^2$ over solid angle is $28/15$.  In linear theory,  $g \approx \Omega_m(z)^{0.6} \, \epsilon^{-1}$.  However, $g$ will depart from the linear theory prediction in part because the absorption saturates in regions.  In the limit that all the absorption is from saturated lines with bias $b$, $g = 1/b$.  Interestingly, \citet{mcdonald03} measured $g=1.5$ at $z=2$, larger than the linear theory prediction of $g\approx 0.5$.  \citet{slosar09} found a value closer to unity from large box, low resolution simulations.  We have also measured $g$ from our $25/h~$Mpc, $2\times 512^3$ hydrodynamic simulation and find values that are consistent with $g \approx 0.5-1$ (Fig. \ref{fig:pkani}), although a larger box size is needed for a more precise determination. 

\section{A Second Derivation of Effect of Intensity Fluctuations on $P_F$}
\label{sec:oldcalc}

This section develops a more sophisticated understanding for how $\Delta_F^2$ is altered by large-scale temperature fluctuations.   
Let us assume $\xi_{Tp7}(r, \Delta_b)$ does not depend on $\Delta_b$, where $\xi_{Tp7}(r, \Delta_b)$ is defined as the correlation function of $T_\Delta^{-0.7}/\langle T_\Delta^{-0.7}\rangle-1$ and $T_\Delta$ is the mean temperature at a given $\Delta_b$. This differs somewhat from our previous definition of the $\delta_{Tp7}$ field as the fluctuation field, but the distinction makes little difference in practice.

We define $\tau_0$ as the Gunn-Peterson optical depth at $\Delta_b= 1$, and the value of $\tau_0$ is set by the mean flux normalization.  
%This effectively determines the value of $T_0^{-0.7} \,\Gamma_{\rm HI}^{-1}$.    
We also define $X \equiv \Delta^{2-.7\gamma}$ and the normalized flux as $F \equiv \exp[-\tau_0 X (1 + \delta_{T})]$.  Finally, we neglect peculiar velocities, and we assume that the temperature fluctuations are uncorrelated with those in density and that they are Gaussian.  While we found in Section \ref{ss:HI_3D} that the correlations with density are important and it is likely that the temperature fluctuations are not Gaussian, the results of this model are illustrative and generalize beyond these assumptions. 

With these assumptions and definitions, the correlation between the normalized flux in a region with density $\Delta_1$ and a region separated by a distance $r$ with $\Delta_2$ is
\begin{eqnarray}
\langle F_1 F_2 \rangle_T &=& \int d \delta_{T_1} d \delta_{T_2} ~e^{-\tau_0 \left(X_1 (1 + \delta_{T_1})+ X_2 (1 + \delta_{T_2}) \right)} \nonumber \\
&  &\times  \, \frac{e^{-\frac{1}{2}(\delta_{T_1} \delta_{T_2}) \bfC^{-1} (\delta_{T_1} \delta_{T_2})^T}}{\sqrt{(2 \pi)^2 \det \bfC}} , \\
&=&  e^{-\tau_0 (X_1+ X_2)} \times  \exp[\frac{\tau_0^2 \sigma_{Tp7}^2}{2} \, (X_1^2 + X_2^2)] \nonumber \\
& & \times  \exp\left[\tau_0^2 \,X_1 X_2  \, \xi_{Tp7}(r)) \right], \label{eqn:prevcorr}\\  
&\approx & \tilde{F}_1 \tilde{F}_2   \left(1 + \tau_0^2 X_1 X_2 \,\xi_{Tp7}(r)\right), \label{eqn:corr}
\end{eqnarray}
where $\tilde{F} \equiv \exp[-\tau_0 X]$, $\bfC \equiv (1,~\xi_{Tp7}/\sigma_{Tp7}^2;~ \xi_{Tp7}/\sigma_{Tp7}^2,~ 1)$, and $\langle ... \rangle_Y$ represents an ensemble average with respect to $Y$.  To go from equation (\ref{eqn:prevcorr}) to equation (\ref{eqn:corr}), we used that $\xi_{Tp7}(r) \ll 1$ to expand the exponential.  Furthermore, we set the second exponential term in equation (\ref{eqn:prevcorr}) to unity.  Since regions that dominate the flux correlation function have $\tau_0 X \sim 1$, the error from dropping this term is of order $\sigma_{Tp7}^2 \ll 1$.  While this is comparable to the temperature term that we kept, it is less interesting because it does not depend on $\xi_{Tp7}(r)$.  In addition, most of its effect is re-absorbed in the renormalization to a single mean flux.  

To calculate the flux correlation function, we average $\langle F_1 F_2 \rangle_T$ over $\Delta_1$ and $\Delta_2$, which yields
\begin{equation}
\xi_F \equiv \langle F_1 F_2 \rangle_{T \Delta} \approx \xi_{\tilde{F}(r)} + K^2 \, \xi_T(r) +  \tau_0^2 \xi_{\tilde{F}}(r) \, \xi_T(r),
\label{eqn:tempmod}
\end{equation}
where $\xi_{\tilde{F}}(r)$ is the unperturbed flux correlation function and
\begin{eqnarray}
K &=& \tau_0 \, \langle F_1 X_1 \rangle_{\Delta_1} , \\
&=&\tau_0 \int d\Delta \, \Delta^{2-.7\gamma} \, \exp[-\tau_0 \Delta^{2-.7\gamma}] \, p
(\Delta).
\end{eqnarray}
The function $p(\Delta_b)$ is the volume-weighted density probability distribution.
We have dropped terms in equation~(\ref{eqn:tempmod}) that involve additional $\xi(r)$ factors as well as the connected moments that are not incorporated in $K^2 \, \xi_T(r)$.

We are interested in the impact of temperature fluctuations on $\gtrsim 10$~Mpc correlations.   At these scales, it is justified to drop the last term in equation (\ref{eqn:tempmod}).  With these simplifications, the power spectrum of fluctuations in the normalized flux is given by
\begin{equation}
P_{F}(k) \approx P_{\tilde{F}}(k) + \frac{K^2}{\bar{F}^2} P_{T}(k).
\label{eqn:3Dpower}
\end{equation}
Thus, the correction proportional to $\xi_T(r)$ has bias $K/\bar{F}$ in this model.  We calculate $K^2/\bar{F}^2 = 0.18$ at $z=4$ and $0.04$ at $z=3$, assuming $T_0 = 2 \times 10^4~$K, $\gamma = 1.3$, and $\Gamma_{\rm HI} = 10^{-12}~$s$^{-1}$, using the \citet{miralda00} fitting function for $p(\Delta)$.   It turns out that $K^2/\bar{F}^2$ is almost identical numerically to the corresponding factor that appears in equation (\ref{eqn:Pklg2}), $b^2/\epsilon^2$.

\end{document}